\documentclass[twocolumn,eqsecnum,floats,aps,nofootinbib,showpacs,preprintnumbers]{revtex4}

\pdfoutput=1 

\usepackage{amsmath,amssymb,amsfonts}
\usepackage{graphicx}

\usepackage{hyperref} 

\def\be{\begin{equation}}
\def\ee{\end{equation}}
\def\ba{\begin{eqnarray}}
\def\ea{\end{eqnarray}}
\def\nn{\nonumber}

\def\a{\alpha}

\def\g{\gamma}
\def\d{\delta}
\def\e{\epsilon}

\def\et{\eta}

\def\k{\kappa}

\def\m{\mu}
\def\n{\nu}
\def\x{\xi}
\def\p{\pi}

\def\r{\rho}
\def\vr{\varrho}
\def\s{\sigma}

\def\t{\tau}

\def\ph{\phi}

\def\G{\Gamma}
\def\D{\Delta}

\newcommand{\abs}[1]{{\left|{#1}\right|}} 

\newcommand{\ftriad}[2]{{}^o\! e^{#1}_{#2}} 
\newcommand{\fcotriad}[2]{{}^o\!\omega_{#1}^{#2}} 
\newcommand{\fq}{{}^o\!q} 
\newcommand{\tE}{\tilde{E}} 

\newcommand{\Pl}{\ell_{\rm Pl}} 
\newcommand{\muzero}{\m^o} 
\newcommand{\Lzero}{{}^o\!L} 
\newcommand{\Vzero}{V_o} 
\newcommand{\azero}{{}^o\!a} 
\newcommand{\mubar}{{\bar \m}} 
\newcommand{\K}{{\cal K}}

\newcommand{\secref}[1]{Section~\ref{#1}}

\newcommand{\eqnref}[1]{(\ref{#1})}
\newcommand{\figref}[1]{FIG.~\ref{#1}}
\newcommand{\appref}[1]{Appendix~\ref{#1}}

\begin{document}

\preprint{IGC-07/9-4}

\title{Effective Dynamics, Big Bounces and Scaling Symmetry\\ in Bianchi Type I Loop Quantum Cosmology}
\author{Dah-Wei Chiou}
\email{chiou@gravity.psu.edu}
\affiliation{
Institute for Gravitation and the Cosmos,
Physics Department,
The Pennsylvania State University, University Park, PA 16802, U.S.A.}

\begin{abstract}
The detailed formulation for loop quantum cosmology (LQC) in the Bianchi I model with a scalar massless field has been constructed. In this paper, its effective dynamics is studied in two improved strategies for implementing the LQC discreteness corrections. Both schemes show that the big bang is replaced by the big bounces, which take place up to three times, once in each diagonal direction, when the area or volume scale factor approaches the critical values in the Planck regime measured by the reference of the scalar field momentum. These two strategies give different evolutions: In one scheme, the effective dynamics is independent of the choice of the finite sized cell prescribed to make Hamiltonian finite; in the other, the effective dynamics reacts to the macroscopic scales introduced by the boundary conditions. Both schemes reveal interesting symmetries of scaling, which are reminiscent of the relational interpretation of quantum mechanics and also suggest that the fundamental spatial scale (area gap) may give rise to a temporal scale.
\end{abstract}

\pacs{04.60.Kz, 04.60.Pp, 98.80.Qc, 03.65.Sq}

\maketitle


\section{Introduction}
The comprehensive formulation for loop quantum cosmology (LQC, see \cite{Bojowald:2006da} for a review) in the spatially flat and isotropic model has been constructed in detail \cite{Ashtekar:2006uz,Ashtekar:2006rx}, giving a conceptual framework and a solid foundation to analyze the physical issues of the quantum theory. With a massless scalar field serving as the \emph{emergent time}, the investigation shows that the quantum evolution is \emph{deterministic across the deep Planck regime} and in the backward evolution of the states which are semiclassical at late times, \emph{the big bang is replaced by a big bounce}.

In the original construction (called ``$\m_o$-scheme'') of \cite{Ashtekar:2006uz,Ashtekar:2006rx}, the discreteness variable introduced to impose the fundamental discreteness of quantum geometry is taken to be constant (referred to as $\m_o$). However, it has been shown that this prescription leads to the wrong semiclassical behavior that, in particular, the critical value of matter density at which the big bounce takes place can be made arbitrarily small by increasing the scalar field momentum $p_\ph$. To fix this problem, the construction was further improved by a more sophisticated implementation of the underlying physical ideas of loop quantum gravity (LQG) with the discreteness variable (referred to as $\mubar$) varying adaptively \cite{Ashtekar:2006wn}. In this improved dynamics (called ``$\mubar$-scheme''), the big bounce occurs precisely when the matter density enters the Planck regime, regardless of the value of $p_\ph$. This construction was also extended to
$k=+1$ and $k=-1$ Robertson-Walker models \cite{Ashtekar:2006es,Vandersloot:2006ws}.

In order to further develop this formulation and extend its domain of validity, based on the same principles of \cite{Ashtekar:2006uz,Ashtekar:2006rx,Ashtekar:2006wn}, both the precursor strategy ($\m_o$-scheme) and the improved strategy ($\mubar$-scheme) were applied and reconstructed for the Bianchi I model to include anisotropy \cite{Chiou:2006qq}. The analytical investigation shows that the state in the kinematical Hilbert space associated with the classical singularity is \emph{completely decoupled} in the difference evolution equation, indicating that the classical singularity is resolved in the quantum evolution and the big bounce may take place when any of the area scales undergoes the vanishing behavior.

While a thorough numerical investigation for \cite{Chiou:2006qq} remains to be done to draw the definite conclusion for the details of the quantum evolution in the Bianchi I model, the effective dynamics with LQC discreteness corrections was first studied in \cite{Date:2005nn} using $\mu_o$-scheme for the vacuum solution and the $\mubar$-scheme effective dynamics has also been done in \cite{Chiou:2007dn} for the case with a massless scalar field and in \cite{Chiou:2007sp} with the inclusion of generic matters. Not only do the results at the level of effective dynamics agree with the anticipations in \cite{Chiou:2006qq} but more intuitive pictures are also obtained in the semiclassical approach, giving insights into how and why the big bounces take place.

Because of the variety of Bianchi I models, extending the improved scheme is slightly ambiguous as a few strategies are possible. Among them, two schemes are of particular interest (referred to as ``$\mubar$-scheme'' and ``$\mubar'$-scheme'' in this paper). The $\mubar$-scheme is suggested in \cite{Chiou:2006qq} and used in \cite{Chiou:2007dn,Chiou:2007sp}, while the $\mubar'$-scheme is disfavored by \cite{Chiou:2006qq} and only mentioned briefly in the appendix of \cite{Chiou:2007sp}. It is later realized that the virtues of $\mubar'$-scheme might have been overlooked and we should not dismiss either of them prematurely in favor of the other before more physics is understood, as either scheme has both advantageous and disadvantageous features.

To better understand the difference of their phenomenological ramifications, this paper presents the details of effective dynamics for the Bianchi I model with a massless scalar field in both $\mubar$- and $\mubar'$-schemes. The investigation shows that in both schemes the big bang singularity is replaced by the big bounces, which take place \emph{up to three times, once in each diagonal direction}. However, $\mubar$- and $\mubar'$-schemes give different evolutions, distinct from each other not only in detail but also qualitatively. For instance, the indication of the occurrence of big bounces is the matter density $\r_\ph$ in $\mubar'$-scheme while it is the ``directional density'' $\vr_I$ in $\mubar$-scheme. Furthermore, in the different schemes, the big bounces conjoin the different pairs (``antipodal'' pair \textit{vs} ``conjugate'' pair ) of classical solutions on the two asymptotic sides of classical regime.

In the Hamiltonian framework, we have to restrict the spatial integration to a finite sized cell ${\cal V}$ to make the Hamiltonian finite. This prescription raises the question whether the resulting dynamics is independent of the choice of ${\cal V}$. This issue is very subtle and has caused a great deal of confusion because the dependence on ${\cal V}$ of the Ashtekar variables is rather obscure. In order to elucidate the subtleties, the classical dynamics is carefully studied both in metric variables and in Ashtekar variables; the precise physical meanings of Ashtekar variables are then obtained. With the subtleties clarified, it can be shown that the effective dynamics in $\mubar'$-scheme is completely independent of the choice of ${\cal V}$ as is the classical dynamics, while the effective dynamics in $\mubar$-scheme in fact reacts to the macroscopic scales introduced by the boundary conditions of ${\cal V}$. This is an important difference between these two schemes.

In addition to the issues related to the dependence on ${\cal V}$, the effective dynamics in both schemes also reveals other interesting symmetries of scaling, which are reminiscent of the relational interpretation of quantum mechanics. It is also suggested that the fundamental scale (area gap) imposed for the spatial geometry may gives rise to a fundamental scale in temporal measurement. These observations could simply be technical artifacts but encourage further investigations for more sophisticated models to tell their profundity.

This paper is organized as follows. In \secref{sec:classical dynamics}, the classical dynamics of the Bianchi I cosmology with a massless scalar source is solved in Hamiltonian formalism in terms of both metric variables and Ashtekar variables. The effective dynamics with LQC discreteness corrections is constructed and solved in \secref{sec:effective dynamics} for the $\mubar$- and $\mubar'$-schemes respectively. The scaling symmetry and issues about relational measurements are discussed in \secref{sec:scaling}. Finally, the results and outlooks are summarized in \secref{sec:summary}. As a comparison to $\mubar$/$\mubar'$-scheme, the effective dynamics in $\m_o$-scheme is also included in \appref{sec:muzero dynamics}. The details of heuristic arguments and motivations for $\mubar$- and $\mubar'$-schemes are given in \appref{sec:mubar schemes}.

\section{Classical Dynamics}\label{sec:classical dynamics}
Before considering the LQC discreteness corrections, in this section, we study the classical dynamics of the Bianchi I cosmology with a massless scalar field. In order to grasp the precise physical meanings of the Ashtekar variables, which are the fundamental degrees of freedom in LQC, the formulations in metric variables (also see \cite{Salisbury:2005sa}) and in Ashtekar variables are both investigated.

\subsection{Formulation in metric variables}
The Hilbert-Einstein action for the cosmology with a massless scalar field $\ph$ minimally coupled to the gravity is given by
\ba\label{eqn:Hilbert action}
S&=&S_{\rm grav}+S_{\ph}:=\int d^4x\,{\cal L}:=\int dt\,L\\
&=&\frac{1}{16\p G}\int d^4x\sqrt{-g}\,R+\frac{1}{2}\int d^4x\sqrt{-g}\,g^{\m\n}\ph_{,\m}\ph_{,\n}.\nn
\ea
In Bianchi I models, the spacetime metric can be written in the diagonal form:
\ba\label{eqn:metric}
ds^2&=&g_{\m\n}dx^\m dx^\n\\
&=&
-dt^2+a_1^2(t)dx^2+a_2^2(t)dy^2+a_3^2(t)dz^2\nn
\ea
and the homogeneous field $\ph({\vec x},t)=\ph(t)$ is independent of the spatial coordinates. The temporal variable $t$ is the proper time and $a_I$ are the diagonal scale factors. The coordinates $(x,y,z)$ which diagonalize $g_{\m\n}$ are the \emph{co-moving coordinates}.

The metric \eqnref{eqn:metric} gives the Ricci scalar:
\be
R=2\left(
\frac{\ddot{a}_1}{a_1}+\frac{\ddot{a}_2}{a_2}+\frac{\ddot{a}_3}{a_3}
+\frac{\dot{a}_2\dot{a}_3}{a_2a_3}+\frac{\dot{a}_1\dot{a}_3}{a_1a_3}
+\frac{\dot{a}_1\dot{a}_2}{a_1a_2}
\right).
\ee
This leads to
\ba
S_{\rm grav}&=&\frac{1}{8\p G}\int d^4x
\biggl\{
\frac{d}{dt}\left(\dot{a}_1a_2a_3+a_1\dot{a}_2a_3+a_1a_2\dot{a}_3\right)\nn\\
&&\qquad\qquad-a_1\dot{a}_2\dot{a}_3-\dot{a}_1a_2\dot{a}_3-\dot{a}_1\dot{a}_2a_3
\biggr\}.
\ea
The total time derivative can be ignored and we then have the Lagrangian
\ba
L&=&\int d^3x
\biggl\{
-\frac{1}{8\p G}
\left(a_1\dot{a}_2\dot{a}_3+\dot{a}_1a_2\dot{a}_3+\dot{a}_1\dot{a}_2a_3\right)\nn\\
&&
\qquad\qquad +a_1a_2a_3 \frac{\dot{\ph}^2}{2}
\biggr\}.
\ea
Because of homogeneity, the spatial integration diverges as we are considering the noncompact Bianchi I model. To circumvent this problem, the integration is restricted to a finite sized cell ${\cal V}$; that is, we take
\be\label{eqn:finite cell}
\int d^3x\longrightarrow \int_{\cal V} d^3x=
\int_0^{L_1}\!\!\!dx \int_0^{L_2}\!\!\!dy \int_0^{L_3}\!\!\!dz
=:V,
\ee
where $L_I$ are the \emph{coordinate} lengths of the edges of ${\cal V}$ and $V$ is its \emph{coordinate} volume.\footnote{\label{footnote:coordinate and physical}The \emph{coordinate} lengths $L_I$ and volume $V$ are independent of time once the cell ${\cal V}$ is chosen since we are working on the \emph{co-moving coordinates}. The \emph{physical} lengths of the edges and volume of ${\cal V}$ are given by ${\bf L}_I:=a_IL_I$ and ${\bf V}:=a_1a_2a_3V$ respectively, which evolve with time.}
[This prescription is equivalent to compactify the space or impose a spatial periodicity. We will see that this does not change the classical dynamics but might have effect on the quantum corrections.]

With this prescription, the Lagrangian is given by
\ba
L&=&
V\biggl\{
-\frac{1}{8\p G}
\left(a_1\dot{a}_2\dot{a}_3+\dot{a}_1a_2\dot{a}_3+\dot{a}_1\dot{a}_2a_3
\right)\nn\\
&&\qquad +a_1a_2a_3 \frac{\dot{\ph}^2}{2}
\biggr\}
\ea
and the canonical momenta of $\ph$ and $a_I$ are defined as:
\ba
\label{eqn:p phi}
p_\ph&:=&\frac{\partial L}{\partial \dot{\ph}}
=Va_1a_2a_3\dot{\ph},\\
\label{eqn:pi}
\p_1&:=&\frac{\partial L}{\partial \dot{a}_1}
=-\frac{V}{8\p G}\left(\dot{a}_2a_3+a_2\dot{a}_3\right)
\ea
and so on for $\p_2$, $\p_3$ in the cyclic manner.\footnote{We will not mention the obvious cyclic repetition hereafter.}
[Note that the definitions of momenta depend on the cell ${\cal V}$ we choose!]
Solving \eqnref{eqn:pi} gives the velocities:
\be
\dot{a}_1=\frac{4\p G}{V}
\left(
\frac{a_1\p_1}{a_2a_3}-\frac{\p_2}{a_3}-\frac{\p_3}{a_2}
\right),
\ee
by which we find the Hamiltonian:
\ba\label{eqn:H}
H&:=&\p_1\dot{a}_1+\p_2\dot{a}_2+\p_2\dot{a}_2+p_\ph\dot{\ph}-L\\
&=&V^{-1}\Biggl\{
2\p G\biggl[-\frac{2\p_2\p_3}{a_1}-\frac{2\p_1\p_3}{a_2}
-\frac{2\p_1\p_2}{a_3}\nn\\
&&\qquad+\frac{a_1\p_1^2}{a_2a_3}+\frac{a_2\p_2^2}{a_1a_3}
+\frac{a_3\p_3^2}{a_1a_2}\biggr]
+\frac{p_\ph^2}{2a_1a_2a_3}
\Biggr\}\nn
\ea
with the canonical relations:
\ba
\{\ph,p_\ph\}&=&1,\\
\label{eqn:a and pi}
\{a_I,\p_J\}&=&\d_{IJ}.
\ea

We can simplify the Hamiltonian by choosing the lapse function $N=Va_1a_2a_3$ and thus introducing the new time variable $dt'=(Va_1a_2a_3)^{-1}dt$.
The rescaled Hamiltonian is then given by
\ba\label{eqn:cl0 rescaled Hamiltonian}
H'&=&2\p G\bigl(-2a_2a_3\p_2\p_3-2a_1a_3\p_1\p_3-2a_1a_2\p_1\p_2\nn\\
&&\qquad+a_1^2\p_1^2+a_2^2\p_2^2+a_3^2\p_3^2\bigr)
+\frac{p_\ph^2}{2}.
\ea
The equations of motion are governed by the Hamilton's equations:
\ba
\label{eqn:cl0 eom 1}
\frac{dp_\ph}{dt'}&=&\{p_\ph,H'\}=0\quad\Rightarrow\
p_\ph\ \text{is constant},\\
\label{eqn:cl0 eom 2}
\frac{d\ph}{dt'}&=&\{\ph,H'\}=p_\ph,\\
\label{eqn:cl0 eom 3}
\frac{da_1}{dt'}&=&\{a_1,H'\}=\frac{\partial\, H'}{\partial \p_1}\nn\\
&=&4\p G\, a_1\left(a_1\p_1-a_2\p_2-a_3\p_3\right),\\
\label{eqn:cl0 eom 4}
\frac{d\p_1}{dt'}&=&\{\p_1,H'\}=-\frac{\partial\, H'}{\partial a_1}\nn\\
&=&-4\p G\, \p_1\left(a_1\p_1-a_2\p_2-a_3\p_3\right).
\ea
In addition, the Hamiltonian must vanish:
\be\label{eqn:cl0 eom 5}
H'(a_I,\p_I,p_\ph)=0.
\ee

Combining \eqnref{eqn:cl0 eom 3} and \eqnref{eqn:cl0 eom 4} gives
\be
\frac{d}{dt'}(a_I\p_I)=0.
\ee
We assign
\be\label{eqn:const Ki}
a_1\p_1=-\hbar(\K_2+\K_3)
\ee
with the dimensionless constants $\K_I$, which will be used to
parameterize the solutions of evolution. Equations \eqnref{eqn:cl0 eom 2} and \eqnref{eqn:cl0 eom 3} then give
\be
\frac{1}{a_I}\frac{da_I}{d\ph}=8\p G\hbar\frac{\K_I}{p_\ph}
=\sqrt{8\p G}\,\frac{\K_I}{\K_\ph},
\ee
if we define
\be\label{eqn:def of p_ph}
p_\ph:=\hbar\sqrt{8\p G}\,\K_\ph
\ee
with the dimensionless constant $\K_\ph$.
Regarding $\ph$ as the \emph{emergent time}, the solutions of evolution are given by
\be\label{eqn:cl sol a}
a_I(\ph)=a_I(\ph_0)\,e^{\sqrt{8\p G}\,\frac{\k_I}{\k_\ph}(\ph-\ph_0)},
\ee
where we scale the parameters $\K_I=\K\k_I$, $\K_\ph=\K\k_\ph$ (with $\K>0$) such that
\be\label{eqn:para constraint 1}
\k_1+\k_2+\k_3=\pm1.
\ee
The Hamiltonian constraint \eqnref{eqn:cl0 eom 5} now reads as
\be\label{eqn:K}
\K_\ph^2=2
\left(\K_2\K_3+\K_1\K_3+\K_1\K_2\right)
\ee
or equivalently
\be\label{eqn:para constraint 2}
\k_1^2+\k_2^2+\k_3^2+\k_\ph^2=1.
\ee

The classical Bianchi I model with a massless scalar field admits both ``Kasner-like''
(two of $\k_I$ positive/negtive and the other negative/positive) and ``Kasner-unlike''
(all $\k_I$ positive/negative) solutions.
The Kasner-like solution, which has two expanding and one contracting directions (say, $\k_\ph>0$ and $\k_1+\k_2+\k_3=1$),
eventually encounters the ``Kasner-like singularity''
(a given regular cubical cell stretches as an infinitely long line) in the far past
and the ``planar collapse'' (a regular cubical cell stretches as an infinitely large plane)
in the far future. On the other hand, the Kasner-unlike solution, with all directions
expanding, encounters the ``Kasner-unlike singularity''
(a regular cubical cell vanishes to a point) in the far past and no planar collapse.\footnote{In the case of no matter sources, the classical dynamics of vacuum Bianchi I model yields the standard \emph{Kasner solution}:
\be
a_I(t)=a_I(t_0)\left(\frac{t}{t_0}\right)^{\k_I},
\ee
where $\k_1,\k_2,\k_3$ are constants subject to:
\be
\k_1+\k_2+\k_3=1\quad\text{and}\quad
\k_1^2+\k_2^2+\k_3^2=1,
\ee
which are to be compared with \eqnref{eqn:para constraint 1} and \eqnref{eqn:para constraint 2}.
Apart from the trivial solution (where two of $\k_I$ vanish), all Kasner solutions must have two of $\k_I$ positive and the other negative.
}

We will see that with LQC discreteness corrections, both Kasner-like and Kasner-unlike singularities are resolved and replaced by the \emph{big bounces},
whereas the planar collapse remains its destiny even one of the three diagonal directions
approaches infinitely small length scale.

Note that, by \eqnref{eqn:p phi} and \eqnref{eqn:pi}, $p_\ph$ and $\p_I$ depend on the choice of ${\cal V}$ and scale as $p_\ph, \p_I \propto V$. Thus, the directly measurable quantities are not the canonical momenta but rather the momentum density of $\ph$:
\be\label{eqn:momentum density}
\frac{p_\ph}{\bf V}\equiv \dot{\ph}={\bf V}^{-1}\hbar\sqrt{8\p G}\,\K_\ph,
\ee
which is nothing but the time derivative of $\ph$,
and the quantities:
\be
\frac{a_1\p_1}{\bf V}\equiv -\frac{1}{2\p G}
\left(H_2+H_3\right)=-\frac{\hbar}{\bf V}\left(\K_2+\K_3\right)
\ee
with the \emph{directional Hubble rates} defined as
\be
H_I:=\frac{\dot{a}_I}{a_I}=\frac{\dot{\bf L}_I}{{\bf L}_I}.
\ee

For given initial physical conditions $\dot{\ph}|_{t_0}$ and $H_I|_{t_0}$, the constants of motion $\K_\ph$ and $\K_I$ both scale as $\propto {\bf V}|_{t_0}$, the physical volume of ${\cal V}$ at the initial time $t_0$. The ratio $\K_I/\K_\ph=\k_I/\k_\ph$ is nevertheless independent of ${\cal V}$ and hence the solution of $a_I(\ph)$ given by \eqnref{eqn:cl sol a} does \emph{not} depend on the choice of ${\cal V}$.
Once $a_I(\ph)$ is solved, to know the solution $a_I(t)$ as a function of $t$, we only need to convert $\ph$ back to $t$ via
\ba\label{eqn:t and phi}
t-t_0&=&\int_{t_0}^t dt=\int_{\ph_0}^\ph\frac{Va_1(\ph)a_2(\ph)a_3(\ph)}{p_\ph}\,d\ph\nn\\
&=&\int_{\ph_0}^\ph\frac{a_1(\ph)a_2(\ph)a_3(\ph)}
{a_1(\ph_0)a_2(\ph_0)a_3(\ph_0)\,\dot{\ph}\bigr|_{t_0}}\,d\ph,
\ea
where, again, the dependence of $V$ is gone.
Therefore, whether in terms of the proper time $t$ or the emergent time $\ph$, the classical dynamics is \emph{completely independent} of the finite sized cell ${\cal V}$ we choose to make sense of the Hamiltonian formalism. [The independence of the choice of ${\cal V}$ is not necessarily retained when quantum corrections are taken into account. In any case, however, the dynamics is independent of ${\cal V}$ in terms of $t$ if and only if it is so in terms of $\ph$.]
Furthermore, \eqnref{eqn:t and phi} gives a simple relation:
\be
t-t_0=\frac{\k_\ph}{\sqrt{8\p G}\,\dot{\ph}\bigr|_{t_0}}
\left(e^{\frac{\sqrt{8\p G}}{\k_\ph}(\ph-\ph_0)}
-1\right)
\ee
for the classical solutions given by \eqnref{eqn:cl sol a}, but the relation between $t$ and $\ph$ is more complicated if the quantum corrections are taken into account.

\subsection{Formulation in Ashtekar variables}
Equivalently, the classical dynamics studied above can be reformulated in terms of Ashtekar variables as the groundwork for LQC formulation. We follow the lines of Section III of \cite{Chiou:2006qq} but emphasize the explicit relations between the metric and Ashtekar variables by rigorously taking care of the subtleties overlooked (and correcting the mistakes made) in \cite{Chiou:2006qq,Chiou:2007dn}.

In Bianchi (homogeneous) models of cosmology, on the homogeneous space-like slice $\Sigma$, we can choose a fiducial triad of vectors $\ftriad{a}{i}$ and a fiducial co-triad of covectors $\fcotriad{a}{i}$ that are left invariant by the action of the Killing fields of $\Sigma$. (Note $\ftriad{a}{i}\fcotriad{b}{i}=\d^a_b$.) The \emph{fiducial} 3-metric of $\Sigma$ is given by the co-triad $\fcotriad{a}{i}$:
\be
\fq_{ab}=\fcotriad{a}{i}\,\fcotriad{b}{j}\,\d_{ij}.
\ee

In LQG, the Ashtekar variables consist of the canonical pairs:
the densitized triads ${\mbox{$\tE$}^a}_i(\vec{x})$ and connections  ${A_a}^i(\vec{x})$, which satisfy the canonical relation:
\be\label{eqn:Poisson of A nd E}
\{{A_a}^i(\vec{x}),{\mbox{$\tE$}^b}_j(\vec{x}')\}
=8\p G\g\,\d^i_j\,\d_a^b\,\d^3(\vec{x}-\vec{x}'),
\ee
where $\g$ is the Barbero-Immirzi parameter.
In the context of Bianchi models, it is more convenient to
consider the \emph{reduced canonical variables}:
\ba
{E^i}_j&:=&\int_{\cal V}d^3x\sqrt{\fq}\left( \fq^{-1/2}\,\fcotriad{a}{i}\,{\mbox{$\tE$}^a}_j\right)\nn\\
&=&\int_{\cal V}d^3x\,\fcotriad{a}{i}\,{\mbox{$\tE$}^a}_j,\\
{A_i}^j&:=&\int_{\cal V}d^3x\sqrt{\fq}\left( {A_a}^j\,\ftriad{a}{i}\right).
\ea
The reduced canonical variables are essentially the same as the original canonical variables, but stripped of the spatial dependence enforced by the Bianchi symmetry and integrated over a finite sized cell ${\cal V}$.\footnote{The spatial integration over a finite sized cell ${\cal V}$ is not explicitly noted in \cite{Chiou:2006qq} and nor in most literatures. This prescription is subtle but important because otherwise the Poisson bracket $\{{E^i}_j, {A_i}^j\}$ would be distributional.}
Conversely, we have the replacement rules:
\ba
\label{eqn:replacement rule1}
{\mbox{$\tE$}^a}_i&\longrightarrow&\Vzero^{-1}\sqrt{\fq}\,{E^j}_i\,\ftriad{a}{j}=
V^{-1}{E^j}_i\,\ftriad{a}{j},\\
\label{eqn:replacement rule2}
{A_a}^i&\longrightarrow&\Vzero^{-1}{A_j}^i\,\fcotriad{a}{j},
\ea
where $\Vzero$ is the \emph{fiducial} volume defined as the volume of ${\cal V}$ measured as if the 3-metric was $\fq_{ab}$, i.e.
\be
\Vzero:=\int_{\cal V} d^3x \sqrt{\fq}\,.
\ee
The canonical relation \eqnref{eqn:Poisson of A nd E} now reads as
\be\label{eqn:Aij and Ekl}
\{{A_i}^j,{E^k}_l\}=8\p G\g \Vzero\d_i^k\d_l^j.
\ee
However, if we compute $\{{A_a}^i,{\mbox{$\tE$}^b}_j\}$ using \eqnref{eqn:replacement rule1} and \eqnref{eqn:replacement rule2} through \eqnref{eqn:Aij and Ekl}, the $\d$-function in the right hand side of \eqnref{eqn:Poisson of A nd E} is correspondingly replaced with
\be
\d^3(\vec{x}-\vec{x}')\longrightarrow V^{-1}
\ee
as expected due to the regularization procedure.

For the diagonal Bianchi class A models (Bianchi I included), ${E^i}_j$ and ${A_i}^j$ are diagonalizable and their diagonalized entries are denoted as $\tilde{p}_I$ and $\tilde{c}_I$ ($I=1,2,3$) respectively.
Equation \eqnref{eqn:Aij and Ekl} then becomes
\be
\{\tilde{c}_I,\tilde{p}_J\}=8\p G\g\Vzero\,\d_{IJ}.
\ee
In Bianchi I models, correspondingly, we choose the fiducial 3-metric to be the diagonal form:
\be
\fq_{ab}=
\left(
\begin{array}{ccc}
\azero_1^2&0&0\\
0&\azero_2^2&0\\
0&0&{\azero_3^2}
\end{array}
\right),
\ee
and the finite sized cell ${\cal V}$ is adapted to be rectangular with respect to $\fq_{ab}$ and coordinated as described in \eqnref{eqn:finite cell}.
The fiducial volume of ${\cal V}$ is now $\Vzero=\Lzero_1\Lzero_2\Lzero_3=\azero_1\azero_2\azero_3 V$ with
\be
\Lzero_I:=\int_0^{L_I} \azero_I dx_I= \azero_I L_I
\ee
being the \emph{fiducial} lengths of the edges of ${\cal V}$.\footnote{The \emph{fiducial} length/volume should not be confused with the \emph{coordinate} or \emph{physical} length/volume (see Footnote~\ref{footnote:coordinate and physical}). (Also note: $\Vzero$ and $\Lzero_I$ could depend on $t$, depending on the choice of $\fq_{ab}$.) Most literatures do not distinguish these notions completely, but the nuances are highlighted in this paper in order to give the precise physical interpretations of $c_I$ and $p_J$.}

To get rid of the nonphysical dependence on $\fq_{ab}$, it is convenient to introduce the new variables:
\ba
p_I&:=&\Lzero_I^{-1}\tilde{p}_I,\\
c_1&:=&(\Lzero_2\Lzero_3)^{-1}\tilde{c}_1,
\ea
which satisfy the canonical relation:
\be\label{eqn:c and p}
\{c_I,p_J\}=8\p G\g\,\d_{IJ}.
\ee
The relation between the densitized triad and the 3-metric is given by
\be\label{eqn:q and E}
qq^{ab}=\d^{ij}{\mbox{$\tE$}^a}_i{\mbox{$\tE$}^b}_j,
\ee
which leads to
\be\label{eqn:p and a}
p_1=L_2L_3(a_2a_3)={\bf L}_2{\bf L}_3.
\ee
Thus, the triad variables $p_I$ are the \emph{physical} areas of the rectangular surfaces of ${\cal V}$.\footnote{More precisely, \eqnref{eqn:p and a} should be $\abs{p_1}={\bf L}_2{\bf L}_3$. The sign of $p_I$ corresponds to spatial orientation, which we ignore in this paper.}
Comparing the canonical relations \eqnref{eqn:a and pi} and \eqnref{eqn:c and p} via \eqnref{eqn:p and a}, we have
\be\label{eqn:c pi a}
c_1=\frac{4\p\g\,G}{L_2L_3a_2a_3} \left(a_1\p_1-a_2\p_2-a_3\p_3\right).
\ee
By \eqnref{eqn:pi}, we have
\be\label{eqn:c and HI}
c_I=\g L_I\dot{a}_I=\g{\bf L}_I H_I=\g \dot{\bf L}_I,
\ee
which tells that, \emph{classically}, the connection variables $c_I$ are the time change rates of the \emph{physical} lengths of the edges of ${\cal V}$ (up to the constant $\g$).
Note that \eqnref{eqn:p and a} and \eqnref{eqn:c pi a} remain the same even with the quantum corrections since the symplectic structures given by \eqnref{eqn:a and pi} and \eqnref{eqn:c and p} as well as the relation \eqnref{eqn:q and E} are unchanged. By contrast, \eqnref{eqn:c and HI} is modified in the presence of quantum effects because Hamiltonian is changed and thus \eqnref{eqn:pi} no longer holds.

By \eqnref{eqn:p and a} and \eqnref{eqn:c pi a}, the classical Hamiltonian \eqnref{eqn:H} can be recast in terms of $c_I$ and $p_I$:
\ba\label{eqn:cl Hamiltonian}
&&H=H_{\rm grav}+H_\ph\\
&=&-\frac{\left(c_2p_2c_3p_3+c_1p_1c_3p_3+c_1p_1c_2p_2\right)}{8\p G\g^2\sqrt{{p_1p_2p_3}}}
+\frac{p_\ph^2}{2\sqrt{{p_1p_2p_3}}}.\nn
\ea
[We can also derive $H_{\rm grav}$ directly from the Hamiltonian constraint of the full theory of LQG. For this approach, see the text toward \eqnref{eqn:H grav} in \appref{sec:mubar schemes} or Section III.C of \cite{Chiou:2006qq}.]
The solutions of $c_I$ and $p_I$ can be readily obtained by translating \eqnref{eqn:const Ki} and \eqnref{eqn:cl sol a} via \eqnref{eqn:p and a} and \eqnref{eqn:c pi a}. But in order to draw a parallel to solving the effective dynamics with LQC corrections studied later, in the following, we present solving the equations of motions directly for $c_I$ and $p_I$.

Again, we simplify the Hamiltonian by choosing the lapse function $N=\sqrt{{p_1p_2p_3}}=Va_1a_2a_3$
and thus introducing the new time variable $dt'=(p_1p_2p_3)^{-1/2}dt=\left(Va_1a_2a_3\right)^{-1}dt$.
The rescaled Hamiltonian (which is the same as \eqnref{eqn:cl0 rescaled Hamiltonian}) is given by
\be\label{eqn:cl rescaled Hamiltonian}
H'=-\frac{\left(c_2p_2c_3p_3+c_1p_1c_3p_3+c_1p_1c_2p_2\right)}{8\p G\g^2}
+\frac{p_\ph^2}{2}.\\
\ee
The equations of motion are governed by \eqnref{eqn:cl0 eom 1} and \eqnref{eqn:cl0 eom 2} as well as
\ba
\label{eqn:cl eom 3}
\frac{dc_1}{dt'}&=&\{c_1,H'\}=8\p G\g\,\frac{\partial\, H'}{\partial p_1}\nn\\
&=&-\g^{-1}c_1\left(c_2p_2+c_3p_3\right),\\
\label{eqn:cl eom 4}
\frac{dp_1}{dt'}&=&\{p_1,H'\}=-8\p G\g\,\frac{\partial\, H'}{\partial c_1}\nn\\
&=&\g^{-1}p_1\left(c_2p_2+c_3p_3\right).
\ea
In addition, the constraint that the Hamiltonian must vanish yields
\ba\label{eqn:cl eom 5}
&&H'(c_I,p_I,p_\ph)=0\quad
\Rightarrow\\
&&p_\ph^2=\frac{1}{4\p G\g^2}
\left(c_2p_2c_3p_3+c_1p_1c_3p_3+c_1p_1c_2p_2\right).\nn
\ea
[Note that substituting \eqnref{eqn:p and a} into \eqnref{eqn:cl eom 4} leads to \eqnref{eqn:c and HI} again.]

Combining \eqnref{eqn:cl eom 3} and \eqnref{eqn:cl eom 4} gives
\ba\label{eqn:const Ki pc}
\frac{d}{dt'}(p_Ic_I)=0,\quad\Rightarrow\quad
p_Ic_I=8\p G\g\hbar\,\K_I,
\ea
where $\K_I$ are the same constants as defined in \eqnref{eqn:const Ki}.
Taking \eqnref{eqn:const Ki pc} into \eqnref{eqn:cl eom 5} yields the same constraint \eqnref{eqn:K}.

Putting \eqnref{eqn:const Ki pc} into \eqnref{eqn:cl eom 4} gives
\be
\frac{1}{p_1}\frac{dp_1}{dt'}={8\p G \hbar}\left(\K_2+\K_3\right).
\ee
Referring to \eqnref{eqn:cl0 eom 2}, this leads to
\be\label{eqn:cl diff eq 2}
\frac{1}{p_1}\frac{dp_1}{d\ph}=8\p G\hbar\,\frac{\K_2+\K_3}{p_\ph}
=\sqrt{8\p G}\left(\frac{1-\k_I}{\k_\ph}\right),
\ee
and consequently
\be\label{eqn:cl sol p}
p_I(\ph)=p_I(\ph_0)\,e^{\sqrt{8\p G}\left(\frac{1-\k_I}{\k_\ph}\right)(\ph-\ph_0)},
\ee
which is the same solution as given by \eqnref{eqn:cl sol a}.

\section{Effective Dynamics with LQC discreteness corrections}\label{sec:effective dynamics}
In LQC, the connection variables $c_I$ do not exist and should be replaced with holonomies. In the effective theory, to capture the quantum corrections, following the procedures used in the isotropic case \cite{Singh:2005xg}, we take the prescription to replace $c_I$ with
\be\label{eqn:c to sin}
c_I\longrightarrow\frac{\sin(\mubar_Ic_I)}{\mubar_I},
\ee
introducing the discreteness variables $\mubar_I$.
The heuristic argument starting from the full theory of LQG and leading to \eqnref{eqn:c to sin} can be found in \appref{sec:mubar schemes}. This prescription can also be understood as the WKB approximation of the full quantum theory of LQC \cite{Date:2005nn}.

Imposing this prescription plus the loop quantum correction to the inverse triad on \eqnref{eqn:cl Hamiltonian}, we have the effective Hamiltonian constraint to the leading order:
\ba\label{eqn:qm Hamiltonian original}
H_{\rm eff}&=&f(p_1)f(p_2)f(p_3)\frac{p_\ph^2}{2}
-\frac{f(p_1)f(p_2)f(p_3)}{8\p G \g^2}\\
&&\quad\times
\left\{
\frac{\sin(\mubar_2c_2)\sin(\mubar_3c_3)}{\mubar_2\mubar_3}p_2p_3+
\text{cyclic terms}
\right\},\nn
\ea
where $f(p_I)$ is the eigenvalue of the inverse triad operator $\widehat{1/\sqrt{{p_I}}}$. The loop quantization gives the quantum corrections on $f(p_I)$:
\be\label{eqn:f(p_I)}
f(p_I)\sim
\left\{
\begin{array}{cr}
\frac{1}{\sqrt{{p_I}}}\left(1+{\cal O}(\Pl^2/p_I)\right) & \text{for}\ p_I\gg\Pl^2\\
\propto{p_I}^n/\Pl^{2n+1} & \text{for}\ p_I\ll\Pl^2
\end{array}
\right.
\ee
with the Planck length $\Pl:=\sqrt{G\hbar}$ and a positive $n$. The corrections to $f(p_I)$ are significant only in the deep Planck regime in the vicinity of $p_I=0$. From now on, we will ignore the quantum corrections to $f(p_I)$ by simply taking its classical function $f(p_I)={p_I}^{-1/2}$. The fact that this effective quantum modification is a good approximation for the sates which are semiclassical at late times has been verified in the isotropic models of LQC sourced with a massless scalar field \cite{Ashtekar:2006wn,Bojowald:2006gr}. The validity of this effective prescription for the vacuum Banchi I model was discussed in the context of WKB approximation in \cite{Date:2005nn} and that for the Bianchi I model with a massless scalar field can also be affirmed \cite{in progress}.
[We will see that in the backward evolution of the solutions which are semiclassical in the late times, the big bounces take place much earlier before the discreteness correction on the inverse triad operator becomes considerable, and it is the ``non-locality'' effect (i.e., using the holonomies) that accounts for the occurrence of big bounces.]

With $f(p_I)={p_I}^{-1/2}$, by choosing $N=\sqrt{{p_ap_2p_3}}$ and $dt'=(p_1p_2p_3)^{-1/2}dt$, the Hamiltonian \eqnref{eqn:qm Hamiltonian original} can be rescaled as
\ba\label{eqn:qm Hamiltonian}
&&H'_\mubar=\frac{p_\ph^2}{2}\\
&&\
-\frac{1}{8\p G \g^2}
\left\{
\frac{\sin(\mubar_2c_2)\sin(\mubar_3c_3)}{\mubar_2\mubar_3}p_2p_3+
\text{cyclic terms}
\right\}.\nn
\ea

As for imposing the fundamental discreteness of LQG on the formulation of LQC, the precursor construction ($\m_o$-scheme) is to take $\mubar_I$ as constants (referred to as $\muzero_I$). However, it has been shown that in the isotropic case $\m_o$-scheme can leads to the wrong semiclassical limit and should be improved by a more sophisticated construction ($\mubar$-scheme) in which the value of $\mubar$ depends adaptively on $p$ (via $\mubar\propto1/\sqrt{{p}}$) and thus implements the underlying physics of quantum geometry of LQG more directly \cite{Ashtekar:2006wn}.
Extending this scheme to the Bianchi I model is slightly ambiguous as a few possibilities exist. Among them, two constructions are most preferable (referred to as ``$\mubar$-scheme'' and ``$\mubar'$-scheme''):
\begin{itemize}
\item
$\mubar$-scheme:
\be\label{eqn:mubar}
\mubar_1=\sqrt{\frac{\D}{{p_1}}}\,,\quad
\mubar_2=\sqrt{\frac{\D}{{p_2}}}\,,\quad
\mubar_3=\sqrt{\frac{\D}{{p_3}}}\,,
\ee
\item
$\mubar'$-scheme:
\be\label{eqn:mubar'}
\mubar'_1=\sqrt{\frac{{p_1}\D}{{p_2p_3}}}\,,\quad
\mubar'_2=\sqrt{\frac{{p_2}\D}{{p_1p_3}}}\,,\quad
\mubar'_3=\sqrt{\frac{{p_3}\D}{{p_1p_2}}}\,.
\ee
\end{itemize}
Here $\D=\frac{\sqrt{3}}{2}(4\p\g\Pl^2)$ is the \emph{area gap} in the full theory of LQG.

Either scheme has its own advantages and drawbacks and until more detailed physics is investigated it is still debatable which one makes better sense. In particular, $\mubar$-scheme is suggested in \cite{Chiou:2006qq}, since in the construction for the full theory of LQC the Hamiltonian constraint in $\mubar$-scheme gives a difference equation in terms of affine variables and therefore the well-developed framework of the spatially flat-isotropic LQC can be straightforwardly adopted. (However, it is argued in \cite{Bojowald:2007ra} that $\mubar$-scheme may lead to an unstable difference equation.) By contrast, $\mubar'$-scheme does not admit the required affine variables and the full LQC of it is very difficult to construct. On the other hand, $\mubar'$-scheme has the virtue that its effective dynamics is independent of the choice of ${\cal V}$ as will be seen. (More detailed comparisons and the heuristic arguments for both schemes are presented in \appref{sec:mubar schemes}.) To explore their differences, we study both $\mubar$-scheme and $\mubar'$-scheme in the context of effective dynamics in \secref{sec:mubar dynamics} and \secref{sec:mubar' dynamics} respectively. (As a comparison, the effective dynamics in $\m_o$-scheme is also presented in \appref{sec:muzero dynamics}, where we see the insensible behavior that $p_I$ can be made arbitrarily small.)

\subsection{Effective dynamics in $\mubar$-scheme}\label{sec:mubar dynamics}
The effective dynamics in $\mubar$-scheme is specified by the Hamiltonian \eqnref{eqn:qm Hamiltonian} with $\mubar_I$ given by \eqnref{eqn:mubar}.
Again, the equations of motion are governed by the Hamilton's equations and the constraint that the Hamiltonian must vanish:
\ba
\label{eqn:qm eom 1}
\frac{dp_\ph}{dt'}&=&\{p_\ph,H'_\mubar\}=0\quad\Rightarrow\
p_\ph\ \text{is constant},\\
\label{eqn:qm eom 2}
\frac{d\ph}{dt'}&=&\{\ph,H'_\mubar\}=p_\ph,
\ea
\ba
\label{eqn:qm eom 3}
\frac{dc_1}{dt'}&=&\{c_1,H'_\mubar\}=8\p G\g\,\frac{\partial\, H'_\mubar}{\partial p_1}\nn\\
&=&-\g^{-1}
\left[\frac{3\sin(\mubar_1c_1)}{2\mubar_1}-\frac{c_1\cos(\mubar_1c_1)}{2}\right]\nn\\
&&\quad\ \times
\left[\frac{\sin(\mubar_2c_2)}{\mubar_2}p_2
+\frac{\sin(\mubar_3c_3)}{\mubar_3}p_3\right],\\
\label{eqn:qm eom 4}
\frac{dp_1}{dt'}&=&\{p_1,H'_\mubar\}=-8\p G\g\,\frac{\partial\, H'_\mubar}{\partial c_1}\nn\\
&=&\g^{-1}p_1\cos(\mubar_1c_1)\nn\\
&&\quad\ \times
\left[\frac{\sin(\mubar_2c_2)}{\mubar_2}p_2
+\frac{\sin(\mubar_3c_3)}{\mubar_3}p_3\right],
\ea
as well as
\ba\label{eqn:qm eom 5}
&&H'_\mubar(c_I,p_I,p_\ph)=0\quad
\Rightarrow\qquad p_\ph^2=\\
&&\quad\frac{1}{4\p G\g^2}
\left\{
\frac{\sin(\mubar_2c_2)\sin(\mubar_3c_3)}{\mubar_2\mubar_3}p_2p_3+
\text{cyclic terms}
\right\}.\nn
\ea
[Note that in the classical limit $\mubar_Ic_I\rightarrow0$, we have
$\sin(\mubar_Ic_I)/\mubar_I\rightarrow c_I$,
$\cos(\mubar_Ic_I)\rightarrow1$ and therefore
\eqnref{eqn:qm eom 3}--\eqnref{eqn:qm eom 5} reduce to their
classical counterparts \eqnref{eqn:cl eom 3}--\eqnref{eqn:cl eom 5}.]
Also notice that substituting \eqnref{eqn:p and a} into \eqnref{eqn:qm eom 4} gives
\ba\label{eqn:qm c and HI}
&&\frac{\sin(\mubar_1c_1)}{\mubar_1}\\
&=&\frac{\g {\bf L}_1}{2}
\left\{
\frac{H_1+H_3}{\cos(\mubar_2c_2)}
+\frac{H_1+H_2}{\cos(\mubar_3c_3)}
-\frac{H_2+H_3}{\cos(\mubar_1c_1)}
\right\},\nn
\ea
which is the modification of \eqnref{eqn:c and HI} with quantum corrections.

Combining \eqnref{eqn:qm eom 3} and \eqnref{eqn:qm eom 4}, we have
\ba\label{eqn:qm dpc/dt'}
&&\left(\frac{3\sin(\mubar_Ic_I)}{2\mubar_I}-\frac{c_I\cos(\mubar_Ic_I)}{2}\right)\frac{dp_I}{dt'}
+p_I\cos(\mubar_Ic_I)\frac{dc_I}{dt'}\nn\\
&&=\frac{d}{dt'}\left[p_I\frac{\sin(\mubar_Ic_I)}{\mubar_I}\right]=0,
\ea
which, in accordance with the classical counterpart \eqnref{eqn:const Ki pc}, yields
\be\label{eqn:qm pc}
p_I\frac{\sin(\mubar_Ic_I)}{\mubar_I}
=8\p G\g\hbar\,\K_I.
\ee\\
Taking \eqnref{eqn:qm pc} into \eqnref{eqn:qm eom 5} again
gives the same constraints on the constant parameters as in
\eqnref{eqn:K}.

Substituting \eqnref{eqn:qm pc} into \eqnref{eqn:qm eom 4} yields
\be\label{eqn:qm diff eq 1}
\frac{1}{p_1}\frac{dp_1}{dt'}
=8\p G \hbar\,\cos(\mubar_1c_1)(\K_2+\K_3).
\ee
By regarding $\ph$ as the emergent time via \eqnref{eqn:qm eom 2} and expressing $\cos x=\pm\sqrt{1-\sin^2 x}$,
\eqnref{eqn:qm diff eq 1} then leads to
\be\label{eqn:qm diff eq 2}
\frac{1}{p_I}\frac{dp_I}{d\ph}=
\pm\sqrt{8\p G}\left(\frac{1-\k_I}{\k_\ph}\right)
\left[1-\frac{\vr_I}{\vr_{I\!,\,{\rm crit}}}\right]^{1/2},
\ee
where we define the \emph{directional density} for the $I$-direction:
\be
\vr_I:=\frac{p_\ph^2}{{p_I}^3}
\ee
and its critical value is given by the \emph{Planckian density}
$\r_{\rm Pl}$ times a numerical factor $\k_\ph^2/\k_I^2$:
\be\label{eqn:qm crit density}
\vr_{I\!,\,{\rm crit}}:=\left(\frac{\k_\ph}{\k_I}\right)^2\r_{\rm Pl},
\qquad
\r_{\rm Pl}:=(8\p G \g^2\D)^{-1}.
\ee
[Note that the directional density $\vr_I$ is of the same dimension as the matter density $\r_\ph:=p_\ph^2/(2{p_1p_2p_3})$ and thus the name.]

From \eqnref{eqn:qm diff eq 2}, we conclude: The big bang singularity (whether Kasner-like or Kasner-unlike) is \emph{replaced by the big bounces}, which take place up to three times, once in each direction whenever each of $\vr_I$ approaches its critical value $\vr_{I,{\rm crit}}$ (and thus $\cos(\mubar_Ic_I)$ flips sign in \eqnref{eqn:qm diff eq 1}).
Furthermore, the differential equations of individual $p_I$ are \emph{completely decoupled} from one another and as a result the epochs of the big bounces in different directions can be arbitrarily separate, depending on the given initial condition. Also notice that the constants $\k_I$ remain the same before and after the bounces. Equivalently, it can be rephrased that the big bounces connect the classical solution specified by $\k_1,\k_2,\k_3$ on one asymptotic side with its ``antipodal'' counterpart on the other asymptotic side specified by the ``antipodal'' parameters:
\be
\k_1,\k_2,\k_3\quad\longleftrightarrow
\quad -\k_1,-\k_2,-\k_3.
\ee
On the other hand, however, the planar collapse in the Kasner-like case is \emph{not} resolved but the vanishing behavior of one of the length scale factors $a_I$ continues.

For given initial conditions, the differential equation \eqnref{eqn:qm diff eq 2} can be solved numerically.\footnote{\label{footnote:initial condition}Equivalently, we can directly solve the coupled differential equations \eqnref{eqn:qm eom 2}--\eqnref{eqn:qm eom 4} for given $p_I(\ph_0)$, $c_I(\ph_0)$ and $p_\ph$. Also note that, to give the initial condition, an alternative way is to specify $p_I(\ph_0)$ together with the constants $\K_I$ and $\K_\ph$. Given $p_I(\ph_0)$ and $\K_I$, $c_I(\ph_0)$ are fixed by \eqnref{eqn:qm pc}.} The evolutions of $p_I(\ph)$, $a_I(\ph)$ and $\vr_I(\ph)$ are depicted in parts (a), (b) and (c) respectively in \figref{fig:fig1} for Kasner-unlike solution and in \figref{fig:fig2} for Kasner-like solution.
Note that the bounces occur at the moments exactly when $\vr_I$ approach their critical values. Also notice that each individual curve of $p_I(\ph)$ and $a_I(\ph)$ in logarithmic scale becomes a straight line before and after the bounce, with the same slope but opposite sign, indicating that the bounces conjoin the two antipodal classical solutions.

It is noteworthy that the directional density $\vr_I$ is the indication of the bounces but the quantity of $\vr_I$ is \emph{not} independent of the choice of ${\cal V}$, as we know
\be
\vr_1=\frac{p_\ph^2}{p_1^3}=\frac{{\bf V}^2\dot{\ph}^2}{{\bf L}_2^2{\bf L}_3^2}=\frac{{\bf L}_1^2}{{\bf L}_2{\bf L}_3}\dot{\ph}^2
\ee
by \eqnref{eqn:momentum density} and \eqnref{eqn:p and a}. Therefore, contrast to the classical dynamics, the effective dynamics in $\mubar$-scheme does \emph{depend} on the choice of the finite sized cell ${\cal V}$. Another subtler dependence on ${\cal V}$ comes from the fact that for given initial physical conditions $\dot{\ph}|_{t_0}$ and $H_I|_{t_0}$, the constant of motion $\K_\ph$ scale as $\propto{\bf V}|_{t_0}$ but $\K_I$ scale as $\propto{\bf V}|_{t_0}$ only approximately (see \eqnref{eqn:qm c and HI} and \eqnref{eqn:qm pc} and notice \eqnref{eqn:qm c and HI} involves the quantum modification with terms $\cos(\mubar_Ic_I)$). As a result, the ratio $\K_\ph/\K_I=\k_\ph/\k_I$ is slightly dependent on ${\cal V}$. (Nevertheless, in the classical regime, we still have $\k_\ph/\k_I\approx(8\p G)^{1/2}\dot{\ph}/H_I$.)

The problem not to be invariant under different choice of ${\cal V}$ is absent in the $\mubar'$-scheme as will be seen in \secref{sec:mubar' dynamics}. However, we should not dismiss $\mubar$-scheme immediately as it is a common phenomenon that a quantum system reacts to macroscopic scales introduced by boundary conditions (for instance,  the well-known ``conformal anomaly'' as a ``soft'' breaking of conformal symmetry). If we have good physical input to tell what exactly the space is to be enclosed as ${\cal V}$ (such as in the compactified Bianchi I model, or applied for the finite sized homogeneous patches in the Belinsky-Khalatnikov-Lifshitz
(BKL) scenario), the dependence on ${\cal V}$ could be rather meritorious than problematic and the bounce occurrence condition ($\vr_I=\vr_{I,{\rm crit}}$) can be understood as: The \emph{physical areas} ($p_1={\bf L}_2{\bf L}_3$, etc) of the surfaces of ${\cal V}$ get bounced when each of them undergoes the Planck regime (times a numerical value $\k_I^2/\k_\ph^2$) measured by the reference of the momentum $p_\ph$.

\subsection{Effective dynamics in $\mubar'$-scheme}\label{sec:mubar' dynamics}
The effective dynamics in $\mubar$-scheme is specified by the Hamiltonian \eqnref{eqn:qm Hamiltonian} with $\mubar_I$ replaced by $\mubar'_I$ given in \eqnref{eqn:mubar'}. For simplicity, we choose a different lapse function $N=(p_1p_2p_3)^{-1/2}$ and correspondingly introduce the new time variable $dt''=(p_1p_2p_3)^{1/2}dt$. With the new lapse, the Hamiltonian \eqnref{eqn:qm Hamiltonian} is further rescaled to the simpler form:
\ba\label{eqn:qm' Hamiltonian}
&&H''_{\mubar'}=\\
&&\quad\r_\ph
-\frac{1}{8\p G \g^2 \D}
\left\{
\sin(\mubar'_2c_2)\sin(\mubar'_3c_3)+
\text{cyclic terms}
\right\},\nn
\ea
where $\r_\ph$ is the matter density of $\ph$:
\be
\r_\ph:=\frac{p_\ph^2}{2{p_1p_2p_3}}:=\frac{p_\ph^2}{2p^3},
\ee
and we define $p^3:=p_1p_2p_3$ for convenience.
Because $\abs{\sin(\mubar'_Ic_I)}\leq 1$, the vanishing of the Hamiltonian constraint $H''_{\mubar'}=0$ immediately implies
\be\label{eqn:rho upper bound}
\r_\ph\leq\frac{3}{8\p G \g^2 \D}=3\r_{\rm Pl}.
\ee
The fact that $\r_\ph$ is bounded above implies that the big-bang singularity is resolved and the big bounce is expected to occur when the matter density approaches Planckian density. The inequality of \eqnref{eqn:rho upper bound} is saturated for the perfectly isotropic universe (i.e. $\k_1=\k_2=\k_3=1/3$) but we shall see that the maximal value of $\r_\ph$ is in general lower than $3\r_{\rm Pl}$ with anisotropies introduced. (Also see Appendix~C of \cite{Chiou:2007sp}.)

To know the detailed dynamics for each individual $p_I$, in addition to the Hamiltonian constraint, we study the Hamilton's equations:
\ba
\label{eqn:qm' eom 1}
\frac{dp_\ph}{dt''}&=&\{p_\ph,H''_{\mubar'}\}=0\quad\Rightarrow\
p_\ph\ \text{is constant},\\
\label{eqn:qm' eom 2}
\frac{d\ph}{dt''}&=&\{\ph,H''_{\mubar'}\}=p_\ph(p_1p_2p_3)^{-1},\\
\label{eqn:qm' eom 3}
\frac{dc_1}{dt''}&=&\{c_1,H''_{\mubar'}\}=8\p G\g\,\frac{\partial\, H''_{\mubar'}}{\partial p_1}\nn\\
&=&-8\p G\g\frac{\r_\ph}{p_1}\\
&&
-\frac{\mubar'_1c_1\cos(\mubar'_1c_1)
\left[\sin(\mubar'_2c_2)+\sin(\mubar'_3c_3)\right]}
{2\g\D\, p_1}\nn\\
&&
+\frac{\mubar'_2c_2\cos(\mubar'_2c_2)
\left[\sin(\mubar'_1c_1)+\sin(\mubar'_3c_3)\right]}
{2\g\D\, p_1}\nn\\
&&
+\frac{\mubar'_3c_3\cos(\mubar'_3c_3)
\left[\sin(\mubar'_1c_1)+\sin(\mubar'_2c_2)\right]}
{2\g\D\, p_1},\nn\\
\label{eqn:qm' eom 4}
\frac{dp_1}{dt''}&=&\{p_1,H''_{\mubar'}\}=-8\p G\g\,\frac{\partial\, H''_{\mubar'}}{\partial c_1}\\
&=&
\frac{\mubar'_1\cos(\mubar'_1c_1)
\left[\sin(\mubar'_2c_2)+\sin(\mubar'_3c_3)\right]}
{\g\D}.\nn
\ea
Note that substituting \eqnref{eqn:p and a} into \eqnref{eqn:qm' eom 4} gives
\ba\label{eqn:qm' c and HI}
&&\frac{\sin(\mubar'_1c_1)}{\mubar'_1}\\
&=&\frac{\g {\bf L}_1}{2}
\left\{
\frac{H_1+H_3}{\cos(\mubar'_2c_2)}
+\frac{H_1+H_2}{\cos(\mubar'_3c_3)}
-\frac{H_2+H_3}{\cos(\mubar'_1c_1)}
\right\},\nn
\ea
which is the modification of \eqnref{eqn:c and HI} with quantum corrections.

From \eqnref{eqn:qm' eom 3} and \eqnref{eqn:qm' eom 4}, we have
\be
2p_1\frac{dc_1}{dt''}=-c_1\frac{dp_1}{dt''}+c_2\frac{dp_2}{dt''}
+c_3\frac{dp_3}{dt''}-16\p G\g\r_\ph,
\ee
which yields
\be\label{eqn:qm' dpc/dt'}
\frac{d}{dt''}\left(p_Ic_I-p_Jc_J\right)=0,
\ee
and hence, in accordance with the constant parameters used for classical solutions in \eqnref{eqn:const Ki pc}, we set
\be\label{eqn:qm' pc}
p_Ic_I
=8\p G\g\hbar\left[\K_I+f(t)\right]
\ee
with the time-varying function $f(t)$. In the classical regime, $p_Ic_I$ become constant and so does $f(t)$. Taking \eqnref{eqn:qm' pc} into the classical Hamiltonian constraint \eqnref{eqn:cl eom 5}, we have
\be
\K_\ph^2=2\Bigl((\K_2+f(t))(\K_3+f(t))+\text{cyclic terms}\Bigr),
\ee
which gives the limiting values for $f(t)$ in the classical regime:
\be
f(t)\rightarrow 0\ \, \text{or}\ -\frac{2}{3}\K\qquad
\text{as}\quad\mubar'_Ic_I\rightarrow0.
\ee
That is, starting with $f=0$ (or $f=-2\K/3$) in the classical regime, $f$ decreases (or increases) towards the big bounces and finally across the bounces $f$ ends up with the other constant $f=-2\K/3$ (or $f=0$) on the other side of the classical regime. Equivalently, it can be said that the big bounces connect the classical solutions specified by $\k_1,\k_2,\k_3$ on one asymptotic side with its ``conjugate'' counterpart on the other asymptotic side specified by the ``conjugate'' parameters:
\be
\k_1,\k_2,\k_3\quad\longleftrightarrow\quad
\k_1-\frac{2}{3},\,\k_2-\frac{2}{3},\,\k_3-\frac{2}{3}.
\ee
[Note that the conjugate parameters also satisfy the constraints \eqnref{eqn:para constraint 1} and \eqnref{eqn:para constraint 2}.]

Substituting \eqnref{eqn:qm' pc} into \eqnref{eqn:qm' Hamiltonian}, $H''_{\mubar'}=0$ then gives the complicated constraint:
\ba\label{eqn:qm' eom 5}
\frac{\ell_\D^6\K_\ph^2}{2p^3}&=&
\sin\left(\frac{\ell_\D^3(\K_2+f)}{p^{3/2}}\right)
\sin\left(\frac{\ell_\D^3(\K_3+f)}{p^{3/2}}\right)\nn\\
&&+\,\text{cyclic terms},
\ea
where we define
\be
\ell_\D^3:=8\p G\g\hbar\,\D^{1/2}\sim\Pl^3
\ee
for convenience.
In particular, considering the classical limit: $f\rightarrow0$ or $f\rightarrow-2\K/3$ and $p^{3/2}:=\sqrt{{p_1p_2p_3}}\gg\ell_\D^3$, we can show that \eqnref{eqn:qm' eom 5} implies \eqnref{eqn:K}.

Substituting \eqnref{eqn:qm' pc} into \eqnref{eqn:qm' eom 4} and regarding $\ph$ as the emergent time via \eqnref{eqn:qm' eom 2}, we have
\ba\label{eqn:qm' diff eq}
\frac{1}{p_1}\frac{dp_1}{d\ph}
\!&=&\!\frac{8\p G \hbar\, p^{3/2}}{p_\ph\,\ell_\D^3}\,
\cos\left(\frac{\ell_\D^3(\K_1\!+\!f)}{p^{3/2}}\right)\\
&&\times\left[
\sin\left(\frac{\ell_\D^3(\K_2\!+\!f)}{p^{3/2}}\right)\!+
\sin\left(\frac{\ell_\D^3(\K_3\!+\!f)}{p^{3/2}}\right)
\right].\nn
\ea
Similar to the case of \eqnref{eqn:qm diff eq 1}, $p_1$ gets bounced once ``$\cos(\cdots)$'' term in \eqnref{eqn:qm' diff eq} flips sign. This happens when
\be\label{eqn:qm' condition 1}
\cos\left(\frac{\ell_\D^3(\K_1\!+\!f)}{p^{3/2}}\right)=0
\quad\Rightarrow\quad
\K_1+f=\frac{\p p^{3/2}}{2\,\ell_\D^3}.
\ee
Assuming $p_2$ and $p_3$ also get bounced roughly around the same moment, at which \eqnref{eqn:qm' condition 1} is satisfied, we have the approximation:
\ba\label{eqn:qm' condition 2}
\sin\left(\frac{\ell_\D^3(\K_2\!+\!f)}{p^{3/2}}\right)
=\sin\left(\frac{\p}{2}+\frac{\ell_\D^3(\K_2-\K_1)}{p^{3/2}}\right)\quad\\
=\cos\left(\frac{\ell_\D^3(\K_2-\K_1)}{p^{3/2}}\right)
\approx 1-\frac{\ell_\D^6(\K_2-\K_1)^2}{2p^3}+\dots\nn
\ea
and similar result for $\sin\left(\ell_\D^3(\K_3+f)/p^{3/2}\right)$.
Substituting \eqnref{eqn:qm' condition 1}, \eqnref{eqn:qm' condition 2} and \eqnref{eqn:K} into \eqnref{eqn:qm' eom 5}, we can solve:
\ba\label{eqn:qm' bounce}
\frac{\ell_\D^3}{p^{3/2}}&\approx&
\sqrt{2}\,\K^{-1}\abs{\k_1-\k_2}^{-1}\abs{\k_1-\k_3}^{-1}\nn\\
&&\times
\Biggl(
(\k_1^2+\k_2^2+\k_3^2)+(\k_1-\k_2)(\k_1-\k_3)\nn\\
&&\quad
-\biggl(
\bigl[
(\k_1^2+\k_2^2+\k_3^2)+(\k_1-\k_2)(\k_1-\k_3)
\bigr]^2\nn\\
&&\qquad\quad-3(\k_1-\k_2)^2(\k_1-\k_3)^2
\biggr)^{1/2}
\Biggr)^{1/2}\nn\\
&=:&\K^{-1} F(\k_1;\k_2,\k_3).
\ea
The approximation is very accurate provided that
\ba
\label{eqn:qm' condition 3}
&&F(\k_1;\k_2,\k_3)\abs{\k_2-\k_1}\ll \p,\nn\\
&&F(\k_1;\k_2,\k_3)\abs{\k_3-\k_1}\ll \p\\
&\Longrightarrow&
\frac{\ell_\D^6(\K_2-\K_1)^2}{p3},\  \frac{\ell_\D^6(\K_3-\K_1)^2}{p3}
\ll \p^2.\qquad
\ea

The criterion \eqnref{eqn:qm' condition 3} is satisfied for the ``near-isotropic'' solutions (i.e. $\k_1$, $\k_2$, $\k_3$ are not so different, or more precisely, $(\k_1-\k_2)^2,\,(\k_1-\k_3)^2\ll\k_1^2+\k_2^2+\k_3^2$). In practice, however, we only need the left hand side of \eqnref{eqn:qm' condition 3} to be fairly smaller than the right hand side in order to have \eqnref{eqn:qm' condition 2} an good approximation, since the Taylor series of $\cos x=1-x^2/2+\cdots$ converges very rapidly.
Therefore, for most solutions (even Kasner-like), we can infer from \eqnref{eqn:qm' bounce} that that big bounces take place up to three times, once in each direction when the matter density $\r_\ph$ approaches one of the three critical values $\r_{I,{\rm crit}}$ given by the Planckian density
$\r_{\rm Pl}:=(8\p G \g^2\D)^{-1}$ times a numerical factor:
\be\label{eqn:qm' crit density}
\r_{1,{\rm crit}}
\approx\frac{1}{2}\k_\ph^2 F^2(\k_1;\k_2,\k_3)\,\r_{\rm Pl}.
\ee
For the solutions far from isotropic and thus \eqnref{eqn:qm' condition 3} is badly violated, the above approximation is no longer good. Nevertheless, the big bounces still take place as most as three times at the moments when the ``$\cos(\cdots)$'' term flips sign in \eqnref{eqn:qm' diff eq}. The matter density $\r_\ph$ is still the indication of the bounce occurrence, although the critical values could be quite different from \eqnref{eqn:qm' crit density}.

The differential equations \eqnref{eqn:qm' eom 2}--\eqnref{eqn:qm' eom 4} can be solved numerically for given initial conditions. [See Footnote~\ref{footnote:initial condition} for specifying the initial condition. Here, given $p_I(\ph_0)$, $\K_I$ and $\K_\ph$, $f(\ph_0)$ can be obtained via \eqnref{eqn:qm' eom 5} and then $c_I(\ph_0)$ are fixed by \eqnref{eqn:qm' pc}.] The evolutions of $p_I(\ph)$, $a_I(\ph)$ and $\r_\ph(\ph)$ are depicted in parts (a), (b) and (c) respectively in \figref{fig:fig3} for Kasner-unlike solution and in \figref{fig:fig4} for Kasner-like solution. Note that in \figref{fig:fig3}, the epochs of big bounces indicated by the dotted lines in (a) are very close to the moments corresponding to $\r_{I, {\rm crit}}$ indicated in (c). In \figref{fig:fig4}, the moments indicated in (a) and those in (c) are quite close to each other but slightly different due to the fact that \eqnref{eqn:qm' condition 3} is now only fairly satisfied. Furthermore, $p_1,p_2,p_3$ bounce roughly around the same time, contrast to the $\mubar$-scheme, in which the epochs of bounces in different directions can be arbitrarily separate. Also notice that each individual curve of $p_I(\ph)$ and $\a_I(\ph)$ in logarithmic scale becomes a straight line before and after the bounce, but the slope is changed, showing that the bounces conjoin two conjugate classical solutions. Finally, as in the $\mubar$-scheme, the planar collapse in the Kasner-like case is \emph{not} resolved.

Contrast to the $\mubar$-scheme dynamics, in which the directional density $\vr_I$ is the indication of bounces, it is the ordinary matter density $\r_\ph$ that signals the occurrence of bounces in $\mubar'$-scheme. Unlike $\vr_I$, the quantity of $\r_\ph$ is independent of the choice of ${\cal V}$ since
\be
\r_\ph=\frac{p_\ph^2}{2{p_1p_2p_3}}
=\frac{{\bf V}^2\dot{\ph}^2}{{\bf V}^2}=\frac{\dot{\ph}^2}{2}.
\ee
Furthermore, \eqnref{eqn:qm' c and HI} implies that the quantity $\mubar'_Ic_I$ depends only on $H_1,H_2,H_3$ and is independent of ${\cal V}$. Consequently, \eqnref{eqn:qm' pc} and \eqnref{eqn:qm' eom 5} tell us: For given initial physical conditions $\dot{\ph}|_{t_0}$ and $H_I|_{t_0}$, $\K_\ph$, $\K_I$ and $f(t)$ all scale as $\propto {\bf V}|_{t_0}$. Therefore, the effective dynamics given by \eqnref{eqn:qm' diff eq} is \emph{completely independent} of the choice of ${\cal V}$ as is the classical dynamics. In particular, the choice of ${\cal V}$ has no effect on the numerical factor $\k_\ph^2 F^2(\k_I;\k_J,\k_K)/2$ appearing in \eqnref{eqn:qm' crit density}. This is a desirable feature which $\mubar$-scheme does not have.
[However, if we further impose the quantum corrections on the eigenvalue of the inverse triad operator as mentioned in \eqnref{eqn:f(p_I)}, this invariance is broken again.]

Even though $\mubar'$-scheme is independent of ${\cal V}$, in case when ${\cal V}$ has a global meaning, the bounce condition ($\r_\ph\approx \r_{I,{\rm crit}}$) can be understood as: The \emph{physical volume} of ${\cal V}$ ($p^{3/2}={\bf V}$) gets bounced when it undergoes the Planck regime (times a numerical value $2\k_\ph^{-2} F^{-2}(\k_I;\k_J,\k_K)$) measured by the reference of the momentum $p_\ph$.

\begin{widetext}

\begin{figure}
\begin{picture}(500,120)(0,0)

\put(5,110){(a)}
\put(175,110){(b)}
\put(345,110){(c)}

\put(0,0)
{
\resizebox{0.98\textwidth}{!}{\includegraphics{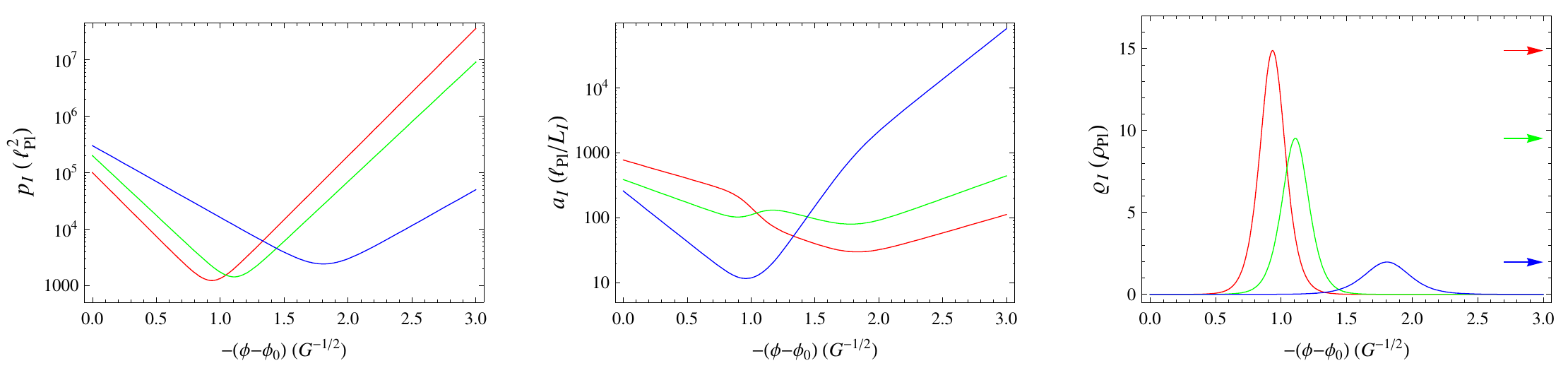}}
}

\end{picture}
\caption{\textbf{Kasner-unlike solution in $\mubar$-scheme effective dynamics.} $\k_1=1/5$, $\k_2=1/4$, $\k_3=11/20$, $\k_\ph=\sqrt{119/200}$; $p_1(\ph_0)=1.\times10^5\Pl^2$,
$p_2(\ph_0)=2.\times10^5\Pl^2$, $p_3(\ph_0)=3.\times10^5\Pl^2$;
and $p_\ph=2.\times10^3\hbar\sqrt{8\p G}$ (i.e., $\K\k_\ph=2.\times10^3$).
The red lines are for $p_1$, $a_1$, $\varrho_1$;
green for $p_2$, $a_2$, $\varrho_2$; and blue for $p_3$, $a_3$, $\varrho_3$. The values of $\varrho_{I,\,{\rm crit}}$ given by \eqnref{eqn:qm crit density} are pointed by the arrows in (c). (The Barbero-Immirzi parameter is set to $\g=1$.)}\label{fig:fig1}
\end{figure}

\begin{figure}
\begin{picture}(500,120)(0,0)

\put(5,110){(a)}
\put(175,110){(b)}
\put(345,110){(c)}

\put(0,0)
{
\resizebox{0.98\textwidth}{!}{\includegraphics{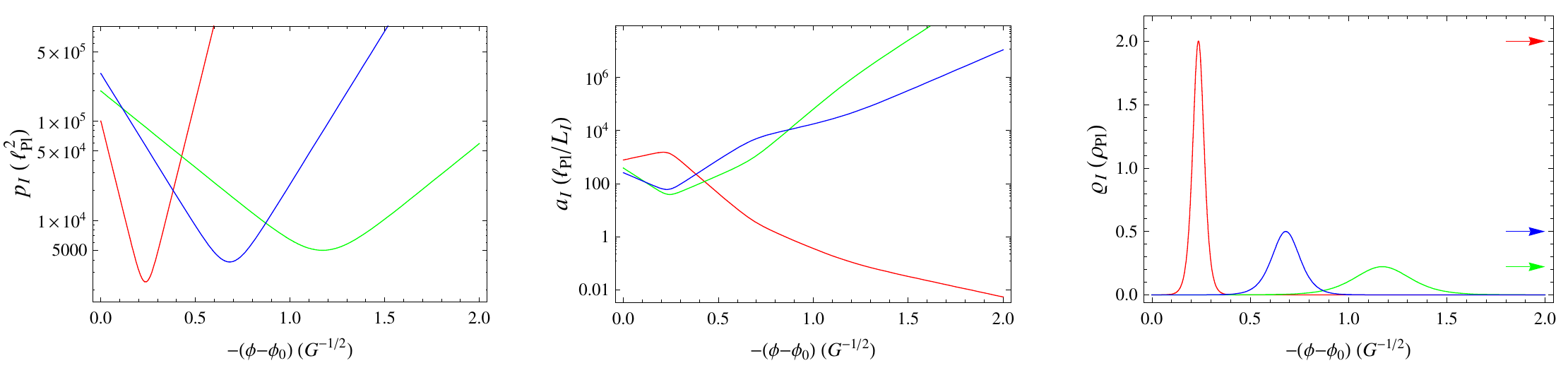}}
}

\end{picture}
\caption{\textbf{Kasner-like solution in $\mubar$-scheme effective dynamics.} $\k_1=-1/4$, $\k_2=3/4$, $\k_3=1/2$, $\k_\ph=1/\sqrt{8}$; $p_1(\ph_0)=1.\times10^5\Pl^2$,
$p_2(\ph_0)=2.\times10^5\Pl^2$, $p_3(\ph_0)=3.\times10^5\Pl^2$;
and $p_\ph=2.\times10^3\hbar\sqrt{8\p G}$ (i.e., $\K\k_\ph=2.\times10^3$).}\label{fig:fig2}
\end{figure}

\begin{figure}
\begin{picture}(500,120)(0,0)

\put(5,110){(a)}
\put(175,110){(b)}
\put(345,110){(c)}

\put(0,0)
{
\resizebox{0.98\textwidth}{!}{\includegraphics{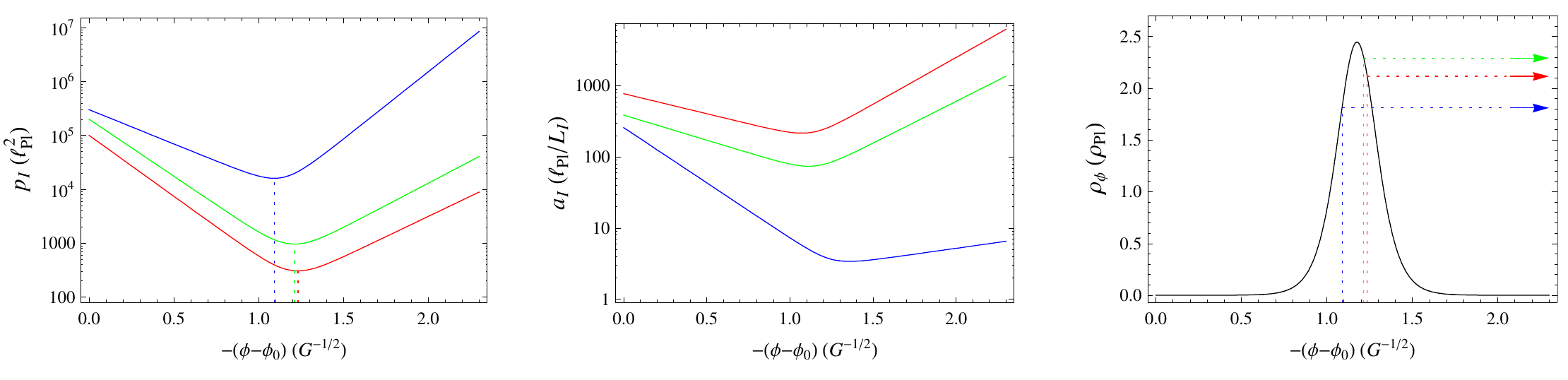}}
}

\end{picture}
\caption{\textbf{Kasner-unlike solution in $\mubar'$-scheme effective dynamics.} $\k_1=1/5$, $\k_2=1/4$, $\k_3=11/20$, $\k_\ph=\sqrt{119/200}$;
$p_1(\ph_0)=1.\times10^5\Pl^2$, $p_2(\ph_0)=2.\times10^5\Pl^2$,
$p_3(\ph_0)=3.\times10^5\Pl^2$;
and $p_\ph=2.\times10^3\hbar\sqrt{8\p G}$ (i.e., $\K\k_\ph=2.\times10^3$).
The values of $\r_{I,{\rm crit}}$ given by \eqnref{eqn:qm' crit density} are pointed by the arrows in (c). The epochs of big bounces are indicated by dotted lines in (a), which are very close to the moments corresponding to $\r_{I,{\rm crit}}$, shown in (c). }\label{fig:fig3}
\end{figure}

\begin{figure}
\begin{picture}(500,120)(0,0)

\put(5,110){(a)}
\put(175,110){(b)}
\put(345,110){(c)}

\put(0,0)
{
\resizebox{0.98\textwidth}{!}{\includegraphics{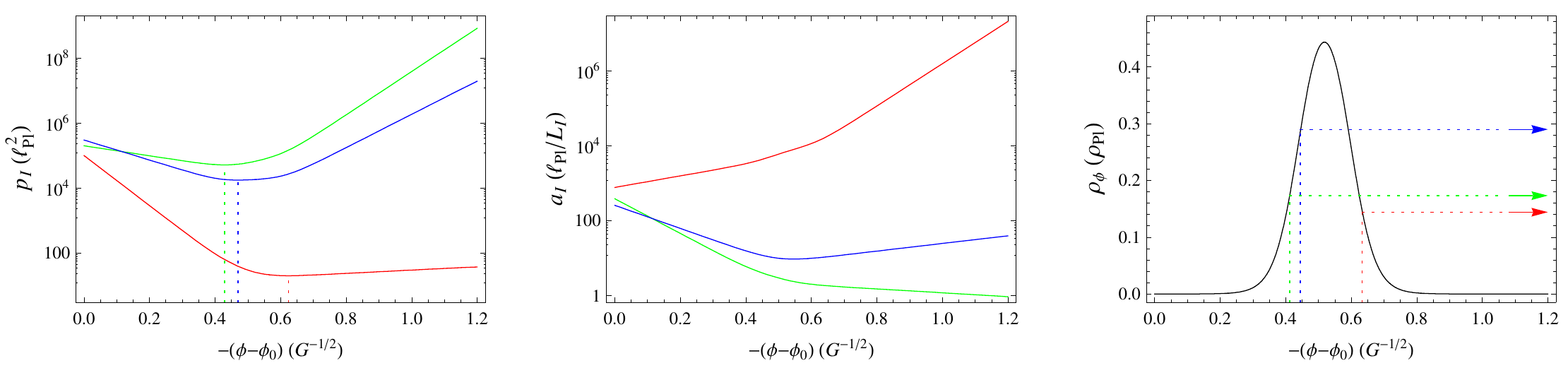}}
}

\end{picture}
\caption{\textbf{Kasner-like solution in $\mubar'$-scheme effective dynamics.} $\k_1=-1/4$, $\k_2=3/4$, $\k_3=1/2$, $\k_\ph=1/\sqrt{8}$; $p_1(\ph_0)=1.\times10^4\Pl^2$,
$p_2(\ph_0)=2.\times10^4\Pl^2$, $p_3(\ph_0)=3.\times10^4\Pl^2$;
and $p_\ph=2.\times10^3\hbar\sqrt{8\p G}$ (i.e., $\K\k_\ph=2.\times10^3$). The epochs of big bounces are only slightly different from the moments corresponding to $\r_{I,{\rm crit}}$.}\label{fig:fig4}
\end{figure}

\end{widetext}

\section{Scaling symmetry and relational measurements}\label{sec:scaling}

With LQC discreteness corrections, both the $\mubar$- and $\mubar'$-schemes show that the big bang singularity (both Kasner-like and Kasner-unlike) is resolved and replaced by the big bounces. The indication of the bounces is the directional densities $\vr_I$ in $\mubar$-scheme whereas it is the matter density $\r_\ph$ in $\mubar'$-scheme.
The detailed evolution of each individual $p_I$ shows that the bounces occur up to three times, once in each direction, when $\vr_I$ in $\mubar$-scheme or $\r_\ph$ in $\mubar'$-scheme approaches its critical value.

On the other hand, the planar collapse is \emph{not} resolved but one of the length scale factors $a_I$ continues the vanishing behavior in the Kasner-like case. This is
expected since the classical solutions \eqnref{eqn:const Ki pc} and \eqnref{eqn:cl sol p} yield $\mubar_Ic_I, \mubar'_Ic_I\rightarrow0$ (and also $\muzero_Ic_I\rightarrow0$ in the $\m_o$-scheme) toward the planar collapse and therefore the quantum corrections become more and more negligible.

The fact that smallness of $p_I$ (not of $a_I$) signals the occurrence of big bounces seems to support the suggestion that ``area is more fundamental than length in LQG'', although whether this is simply a technical artifact or reflects some deep physics is still not clear. (See Section VII.B of \cite{Rovelli:1997yv} for some comments on this aspect and \cite{Rovelli:1993vu} for more details.)
Meanwhile, as the length operator has been shown to have a
discrete spectrum \cite{Thiemann:1996at}, the fact that the
vanishing of the length scale factor in the planar collapse is not stopped seems to contradict the discreteness of the length spectrum. Whether we miss some important ingredients when imposing the fundamental discreteness of LQG in the LQC construction or indeed area is more essential than length remains an open question for further investigation.

It has also been noted that the classical dynamics and the effective dynamics in $\mubar'$-scheme are both independent of the choice of the finite sized cell ${\cal V}$, while the effective dynamics in $\mubar$-scheme depends on the physical size of ${\cal V}$. This can be rephrased in terms of the scaling symmetry;\footnote{A dynamical system is said to be invariant under a certain scaling if for a given solution ($p_I(t)$, $c_I(t)$, $\ph(t)$ and $p_\ph$) to the dynamics, the rescaled functions also satisfy the equations of motion (i.e. Hamilton's equations and vanishing of Hamiltonian constraint). For classical dynamics, the equations to be satisfied are \eqnref{eqn:cl0 eom 1}, \eqnref{eqn:cl0 eom 2} and \eqnref{eqn:cl eom 3}--\eqnref{eqn:cl eom 5}; for $\mubar$-scheme, \eqnref{eqn:qm eom 1}--\eqnref{eqn:qm eom 5}; and for $\mubar'$-scheme, \eqnref{eqn:qm' eom 1}--\eqnref{eqn:qm' eom 4} and \eqnref{eqn:qm' eom 5}.} that is, the classical dynamics and $\mubar'$-scheme effective dynamics are invariant under the following scaling:
\ba
p_1,p_2,p_3&\longrightarrow& l_2l_3 p_1,\,l_1l_3 p_2,\,l_1l_2 p_3,\nn\\
c_1,c_2,c_3&\longrightarrow& l_1 c_1,\,l_2 c_2,\,l_3 c_3,\nn\\
p_\ph&\longrightarrow&l_1l_2l_3 p_\ph,\nn\\
\K_\ph&\longrightarrow&l_1l_2l_3 \K_\ph,\nn\\
\K_I&\longrightarrow&l_1l_2l_3 \K_I.
\ea
[Note that the scaling for $\K_I$ should be accompanied by the same scaling on $f$ in $\mubar'$-scheme; in this case, that is $f\longrightarrow l_1l_2l_3 f$.]
By contrast, the $\mubar$-scheme does not respect this scaling except for the special case with $l_1=l_2=l_3$. In this sense, $\mubar$-scheme has less symmetry of scaling than $\mubar'$-scheme and the classical dynamics.

Apart from the scaling related to the dependence on ${\cal V}$, all three theories (classical, $\mubar$-scheme, $\mubar'$-scheme) also admit the scaling symmetry under:
\ba
\ph&\longrightarrow&\lambda\ph,\nn\\
p_1,p_2,p_3&\longrightarrow&
\lambda^{2/3}p_1,\,\lambda^{2/3}p_2,\,\lambda^{2/3} p_3,\nn\\
c_1,c_2,c_3&\longrightarrow& \lambda^{1/3}c_1,\,\lambda^{1/3}c_2,\,\lambda^{1/3}c_3,\nn\\
p_\ph&\longrightarrow&\lambda p_\ph,\nn\\
\K_\ph&\longrightarrow&\lambda \K_\ph,\nn\\
\K_I&\longrightarrow&\lambda \K_I.
\ea
This is reminiscent of the idea as suggested in \cite{Rovelli:1990ph,Rovelli:1992vv} that length/area/volume is measurable only if the line/surface/bulk is \emph{coupled with the material reference}.

Another symmetry of scaling respected by the classical dynamics is given by
\ba
t&\longrightarrow&\et t,\nn\\
\g&\longrightarrow&\x \g,\nn\\
c_1,c_2,c_3&\longrightarrow& \x\et^{-1}c_1,\,\x\et^{-1}c_2,\,\x\et^{-1}c_3,\nn\\
p_\ph&\longrightarrow&\et^{-1} p_\ph,\nn\\
\K_\ph&\longrightarrow& \et^{-1} \K_\ph,\nn\\
\K_I&\longrightarrow&\et^{-1} \K_I.
\ea
The scaling symmetry regarding $\g\longrightarrow \x \g$ is expected, since the Barbero-Immirzi parameter $\g$ has no effect on the classical dynamics. The scaling symmetry regarding $t\longrightarrow \et t$ is also easy to understand, since there is no temporal scale introduced in the Hamiltonian. However, very surprisingly, the scaling symmetry involving $t\longrightarrow \et t$ is violated for both $\mubar$-scheme and $\mubar'$-scheme effective dynamics. Curiously, this symmetry is restored if $t\longrightarrow \et t$ is accompanied by $\g \longrightarrow\x \g$ and one extra scaling is also imposed at the same time:
\be
\D\longrightarrow \x^{-2}\et^2\D.
\ee
This intriguing observation seems to suggest, albeit speculatively, that in the context of quantum gravity the fundamental scale (area gap) in spatial geometry gives rise to a temporal scale via the non-locality of quantum gravity (i.e., using holonomies) and the Barbero-Immirzi parameter $\g$ somehow plays the role bridging the scalings in time and space. [This reminds us that, in LQG, the precise value of the area gap $\D$ is proportional to $\g$, and $\g$ is also the parameter which relates the \emph{intrinsic} geometry (encoded by spin connection $\G^i_a$) with the \emph{extrinsic} curvature (${K_a}^i$) via ${A_a}^i=\G^i_a-\g{K_a}^i$.]

Meanwhile, related to the above observations, the physical meaning of the directional factors $\vr_I$ and matter density $\r_\ph$ can be interpreted as the (inverse of) area and volume scales, \emph{measured by the reference of the matter content}. In this regard, we may say that the big bounces take place when one of the area scales or the volume scale becomes very small by the reference of the matter momentum.
It is then tempting to regard not only $\ph$ as the ``internal clock'' (emergent time) but also $p_\ph$ as the ``internal rod'' --- namely, the measurement of both temporal and spatial geometries makes sense only in the presence of matter content.

The observations about the symmetry of the scalings $\ph\longrightarrow \lambda \ph$, $t\longrightarrow\et t$ and $\g\longrightarrow\x\g$ may support the ideas of the relational interpretation of quantum mechanics with real rods and clocks such as studied in \cite{Gambini:2006ph} (see also \cite{Rovelli:1990ph,Rovelli:1992vv}), although the link is far from clear. However, as the theories studied in this paper are highly simplified, these observations could be just artifacts and we should not take this speculation too seriously until this aspect is further investigated for more sophisticated models.

Unfortunately, all the scaling symmetries break down in the detailed construction of LQC with $\mubar$-scheme (the strategy to construct the full theory of LQC in $\mubar'$-scheme is still not clear) even for the isotropic model (where $\mubar$- and $\mubar'$-schemes are identical). The full quantum theory only respects the scaling symmetries at the leading order. This is due to the fact that the quantum evolution in the full theory of LQC is governed by a difference equation, in which the step size of difference introduces an additional scale in the deep Planck regime (see \cite{Chiou:2006qq} for Bianchi I model and \cite{Ashtekar:2006wn} for the isotropic model). In fact, already in the level of effective dynamics, the scaling symmetries are violated if we further take into account the LQC corrections on the inverse triad as given by \eqnref{eqn:f(p_I)}. For the full theory of LQC, if we take the aforementioned symmetries seriously, we might be able to revise the detailed construction in the spirit of relational quantum theory such that the step size in the difference equation scales adaptively by the reference of the matter content.

\begin{widetext}


\begin{table}
\begin{tabular}{|c|c|c|}
\hline\hline
\textbf{classical dynamics} & \textbf{effective dynamics in $\mubar$-scheme} & \textbf{effective dynamics in $\mubar'$-scheme}\\

\hline\hline
$p_\ph=\hbar\sqrt{8\p G}\K_\ph={\bf V}\dot{\ph}$ & $p_\ph=\hbar\sqrt{8\p G}\K_\ph={\bf V}\dot{\ph}$ & $p_\ph=\hbar\sqrt{8\p G}\K_\ph={\bf V}\dot{\ph}$\\

\hline
$p_1=L_2L_3(a_2a_3)={\bf L}_2{\bf L}_3$ & $p_1=L_2L_3(a_2a_3)={\bf L}_2{\bf L}_3$ & $p_1=L_2L_3(a_2a_3)={\bf L}_2{\bf L}_3$\\

\hline
$c_I=
\g{\bf L}_I H_I=\g \dot{\bf L}_I$ &
\begin{tabular}{l}
$\frac{\sin(\mubar_1c_1)}{\mubar_1}=$\\
$\frac{\g {\bf L}_1}{2}
\left\{
\frac{H_1+H_3}{\cos(\mubar_2c_2)}
+\frac{H_1+H_2}{\cos(\mubar_3c_3)}
-\frac{H_2+H_3}{\cos(\mubar_1c_1)}
\right\}$
\end{tabular}
&
\begin{tabular}{l}
$\frac{\sin(\mubar'_1c_1)}{\mubar'_1}=$\\
$\frac{\g {\bf L}_1}{2}
\left\{
\frac{H_1+H_3}{\cos(\mubar'_2c_2)}
+\frac{H_1+H_2}{\cos(\mubar'_3c_3)}
-\frac{H_2+H_3}{\cos(\mubar'_1c_1)}
\right\}$
\end{tabular}\\

\hline
$p_Ic_I=8\p G\g\hbar\,\K_I=\g{\bf V}H_I$ & $p_I\frac{\sin(\mubar_Ic_I)}{\mubar_I}
=8\p G\g\hbar\,\K_I$ &
\begin{tabular}{c}
$p_Ic_I
=8\p G\g\hbar\left[\K_I+f(t)\right]$\\
$f\rightarrow 0,-\frac{2}{3}\K$\quad in classical regime
\end{tabular}\\

\hline
$\K_\ph^2=2\left(\K_2\K_3+\K_1\K_3+\K_1\K_2\right)$ & $\K_\ph^2=2\left(\K_2\K_3+\K_1\K_3+\K_1\K_2\right)$ &
\begin{tabular}{c}
$\K_\ph^2=2\left(\K_2\K_3+\K_1\K_3+\K_1\K_2\right)$\quad and\\
$\K_\ph^2=2\biggl\{
\frac{p^3}{\ell_\D^6}
\sin\left(\frac{\ell_\D^3(\K_2+f)}{p^{3/2}}\right)
\sin\left(\frac{\ell_\D^3(\K_3+f)}{p^{3/2}}\right)$\\
$+$\,cyclic terms$\biggr\}$
\end{tabular}\\

\hline
$\frac{1}{p_1}\frac{dp_1}{d\ph}=\frac{8\p G\hbar}{p_\ph}\left(\K_2+\K_3\right)$
&
$\frac{1}{p_1}\frac{dp_1}{d\ph}=\frac{8\p G\hbar}{p_\ph}\cos(\mubar_1c_1)\left(\K_2+\K_3\right)$
&
\begin{tabular}{l}
$\frac{1}{p_1}\frac{dp_1}{d\ph}
=\frac{8\p G \hbar}{p_\ph}\,
\cos\left(\frac{\ell_\D^3(\K_1+f)}{p^{3/2}}\right)$\\
\qquad$\times\frac{p^{3/2}}{\ell_\D^3}\left\{
\sin\left(\frac{\ell_\D^3(\K_2+f)}{p^{3/2}}\right)+
\sin\left(\frac{\ell_\D^3(\K_3+f)}{p^{3/2}}\right)
\right\}$
\end{tabular}\\

\hline
\begin{tabular}{c}
$p_I\rightarrow 0$\\
toward singularity\\
at the same moment\\
$\mbox{}$
\end{tabular} &
\begin{tabular}{l}
$p_I$ bounces whenever\\
$\vr_I:=\frac{p_\ph^2}{{p_I}^3}
=\frac{\k_\ph^2}{\k_I^2}\r_{\rm Pl}$;\\
epochs of the bounces can\\
be arbitrarily separate
\end{tabular} &
\begin{tabular}{l}
$p_1$ bounces at the moment when\\
$\r_\ph:=\frac{p_\ph^2}{2{p_1p_2p_3}}\approx
\frac{1}{2}\k_\ph^2 F^2(\k_1;\k_2,\k_3)\,\r_{\rm Pl}$;\\
$p_1,p_2,p_3$ bounce
roughly around\\
the same moment
\end{tabular}\\

\hline
\begin{tabular}{c}
no big bounce;\\
$\k_I$ fixed
\end{tabular} &
\begin{tabular}{c}
the big bounces conjoin the pair\\
of ``antipodal'' classical solutions:\\
$\k_I\longleftrightarrow-\k_I$
\end{tabular}
&
\begin{tabular}{c}
the big bounces conjoin the pair\\
of ``conjugate'' classical solutions:\\
$\k_I\longleftrightarrow \k_I-\frac{2}{3}$
\end{tabular}\\

\hline
\begin{tabular}{c}
symmetry of scaling:\\
$t\longrightarrow\et t$\\
$\g\longrightarrow\x \g$\\
$\ph\longrightarrow
\lambda\ph$\\
$p_1,p_2,p_3\longrightarrow l_2l_3\lambda^{2/3} p_1,\ \dots$\\
$c_1,c_2,c_3\longrightarrow l_1\lambda^{1/3}\x\et^{-1}c_1,\ \dots$\\
$p_\ph\longrightarrow
l_1l_2l_3\lambda \et^{-1}p_\ph$\\
$\K_\ph\longrightarrow
l_1l_2l_3\lambda \et^{-1}\K_\ph$\\
$\K_I\longrightarrow
l_1l_2l_3\lambda \et^{-1}\K_I$\\
$\mbox{}$
\end{tabular}&
\begin{tabular}{c}
symmetry of scaling:\\
$t\longrightarrow\et t$\\
$\g\longrightarrow\x \g$\\
$\ph\longrightarrow
\lambda\ph$\\
$p_1,p_2,p_3\longrightarrow l^2\lambda^{2/3} p_1,
\ \dots$\\
$c_1,c_2,c_3\longrightarrow l\lambda^{1/3}\x\et^{-1}c_1,
\ \dots$\\
$p_\ph\longrightarrow
l^3\lambda \et^{-1}p_\ph$\\
$\K_\ph\longrightarrow
l^3\lambda \et^{-1}\K_\ph$\\
$\K_I\longrightarrow
l^3\lambda \et^{-1}\K_I$\\
$\D\longrightarrow\x^{-2}\et^2\D$
\end{tabular}&
\begin{tabular}{c}
symmetry of scaling:\\
$t\longrightarrow\et t$\\
$\g\longrightarrow\x \g$\\
$\ph\longrightarrow
\lambda\ph$\\
$p_1,p_2,p_3\longrightarrow l_2l_3\lambda^{2/3}p_1,
\ \dots$\\
$c_1,c_2,c_3\longrightarrow l_1\lambda^{1/3}\x\et^{-1}c_1,
\ \dots$\\
$p_\ph\longrightarrow
l_1l_2l_3\lambda \et^{-1}p_\ph$\\
$\K_\ph\longrightarrow
l_1l_2l_3\lambda\et^{-1}\K_\ph$\\
$(\K_I+f)\longrightarrow
l_1l_2l_3\lambda\et^{-1}(\K_I+f)$\\
$\D\longrightarrow\x^{-2}\et^2\D$
\end{tabular}\\

\hline\hline
\end{tabular}
\caption{Summary of the classical dynamics, $\mubar$-scheme effective dynamics and $\mubar'$-scheme effective dynamics.}\label{tab:summary}
\end{table}

\end{widetext}

\section{Summary and outlook}\label{sec:summary}

To summarize, we list the important facts for the classical dynamics, $\mubar$-scheme effective dynamics and $\mubar'$-scheme effective dynamics in TABLE~\ref{tab:summary}. In the following, the main results are restated and some feasible extensions are remarked.

With the LQC discreteness corrections, the effective dynamics shows that the classical singularities (both Kasner-like and Kasner-unlike) are resolved and \emph{replaced by the big bounces}, which take place \emph{up to three times}, once in each diagonal direction. In $\mubar$-scheme, it is the directional densities $\vr_I$ that signal the occurrence of big bounces whenever each of $\vr_I$ approaches its critical value. In $\mubar'$-scheme, the indication is the matter density $\r_\ph$ and the big bounces happen at the moments when $\r_\ph$ approaches one of the three critical values.

Furthermore, the detailed evolution shows that the equations of motion (in terms of emergent time $\ph$) for $p_I$ in different diagonal directions are completely decoupled and evolve independently of one another in $\mubar$-scheme; as a result the moments of three bounces can be arbitrarily separate. By contrast, in $\mubar'$-scheme, equations of motion for $p_I$ are coupled through $\r_\ph$ and thus $p_1$, $p_2$, and $p_3$ bounce roughly around the same moment. In both schemes, there are three bounces in general unless the initial conditions are delicately fine tuned.

On the other hand, the planar collapse is \emph{not} resolved but one of the length scale factors $a_I$ continues the vanishing behavior in the Kasner-like case. Whether this suggests that area is more fundamental than length or conflicts with the discrete spectrum of length operator in LQG requires further investigation.

Across the bounces, the equations of motion again come closer and closer to the classical ones. Hence, the semiclassicality is retained on both asymptotic sides of the evolution. Given a classical solution in one asymptotic side, the evolution ends up with another classical solution on the other asymptotic side, which is either \emph{antipodal} or \emph{conjugate} to the given solution in $\mubar$/$\mubar'$-scheme respectively.

In regard to the finite sized cell ${\cal V}$ chosen to make sense of the Hamiltonian formalism, the effective dynamics in $\mubar$-scheme depends on the choice of ${\cal V}$, and thereby reacts to the macroscopic scales introduced by the boundary conditions. (In terms of symmetry, it is said that $\mubar$-scheme has less scaling symmetry than $\mubar'$-scheme and classical dynamics.) The effective dynamics in $\mubar'$-scheme, by contrast, is completely independent of ${\cal V}$ as is the classical dynamics. (The issue of dependence on ${\cal V}$ may be related to the instability of $\mubar$-scheme indicated in \cite{Bojowald:2007ra}.) In case that the physical size of ${\cal V}$ has a global meaning (such as in the compactified Bianchi I model or the finite sized homogeneous patches in BKL scenario), the condition for the bounce occurrence can be rephrased: In $\mubar$/$\mubar'$-scheme (respectively), the physical \emph{area/volume} of the surfaces/bulk of ${\cal V}$ gets bounced when it undergoes the Planck regime (times a numerical value) measured by the reference of the momentum $p_\ph$.

While the $\mubar'$-scheme has the advantage that its effective dynamics is independent of ${\cal V}$, the full theory of LQC based on $\mubar'$-scheme is difficult to construct. Both $\mubar$- and $\mubar'$-schemes have desirable merits and it is still disputable which one is more faithful implementation of the underlying physics of quantum geometry.

In addition to the symmetry related to the choice of ${\cal V}$, both schemes admit additional symmetries of scaling, which are reminiscent of the relational interpretation of quantum mechanics, featuring the ideas of real rods and clocks. Furthermore, the symmetry involving the Barbero-Immirzi parameter is suggestive that the fundamental scale (area gap) in spatial geometry may give rise to a fundamental scale in temporal measurement. These symmetries however break down in the construction for the full theory of LQC.

A few possible extensions following the treatment of this paper seem to be within reach. First, as this paper focuses specifically on the model with a massless scalar field, it should be straightforward (with necessary approximation) to extend the results to the models with inclusion of generic matters, which will fulfill the investigation in \cite{Chiou:2007sp} for the $\mubar'$-scheme. Another extension involves the inclusion of nontrivial potentials for the scalar field. In both extensions, most of the observations we obtained here are to be expected and the investigations in broader context would further support or oppose some of our speculations.

Second, it would be instructive to compare the results of this paper with the perturbative treatment of anisotropies, since it has been suggested that the big bang singularity is not resolved if anisotropies are treated as perturbations of an isotropic background \cite{Bojowald:2005wh}. Incorporation of inhomogeneities, on the other hand, will be substantially more difficult. In order to determine whether $\mubar$- or $\mubar'$-scheme is more physically sensible, we have to incorporate inhomogeneities at least approximately or perturbatively in the background of Bianchi I type. The perturbative treatment such as developed in \cite{Bojowald:2006qu} or the strong gravity approximation proposed in \cite{strong gravity} might be useful for this purpose.

Finally, the results in principle could be extended to the Bianchi IX model. Since Bianchi IX model, along with Bianchi I model, plays a pivotal role for the infinite fragmentation of the homogeneous patches close to the singularity in the BKL scenario, the analysis may provide hints to how the loop quantum effects resolve the classical singularities in generic situations and in return it may help arbitrate the tension between $\mubar$- and $\mubar'$-schemes for the Bianchi I model.


\begin{acknowledgements}
The author is grateful for the valuable discussions with Abhay Ashtekar, Martin Bojowald, Golam Hossain, Kevin Vandersloot and especially Tomasz Pawlowski and Parampreet Singh, who helped clarify confusions and inspired important ideas. This work was supported in part by the NSF grant PHY-0456913.
\end{acknowledgements}

\appendix

\section{Effective Dynamics in $\m_o$-Scheme}\label{sec:muzero dynamics}
One of the virtues of the improved strategy ($\mubar$-scheme) in the isotropic model is to fix the serious drawback in the old precursor strategy ($\m_o$-scheme) that the critical value of the matter density at which the bounce occurs can be made arbitrarily small by increasing the momentum $p_\ph$, thereby leading to wrong semiclassical behavior.

In the Bianchi I model, either the directional densities $\vr_I$ (in $\mubar$-scheme) or the matter density $\r_\ph$ (in $\mubar'$-scheme) plays the same role as $\r_\ph$ does in the isotropic case. Having learned from the isotropic case, we expect that the critical values of $\vr_I$ or $\r_\ph$ at which the bounces occur can be made arbitrarily small by increasing the momentum $p_\ph$ in $\m_0$-scheme but is independent of $p_\ph$ in $\mubar$- or $\mubar'$-scheme. The latter is what has been shown in the main text of this paper.\footnote{The critical values $\vr_{I,{\rm crit}}$  depend on $p_\ph$ only through the ratios $\k_\ph^2/\k_I^2$ in $\mubar$-scheme; $\r_{I,{\rm crit}}$ depends on $p_\ph$ only through $\k_\ph^2F^2(\k_I;\k_J,\k_K)/2$ in $\mubar'$-scheme.} For comparison, the effective dynamics in $\m_o$-scheme is presented here.\footnote{For the Bianchi I LQC, the effective dynamics in $\m_o$-scheme was first studied in \cite{Date:2005nn} for the vacuum solution.}

In the effective theory of $\m_o$-scheme, we take the
prescription to replace $c_I$ by $\sin(\muzero_Ic_I)/\muzero_I$ with the \emph{fixed} numbers $\muzero_I$ for discreteness. Analogous to \eqnref{eqn:qm Hamiltonian},
we have the effective (rescaled) Hamiltonian constraint:
\ba
&&H'_{\m_o}=\frac{p_\ph^2}{2}\\
&&\
-\frac{1}{8\p G \g^2}
\left\{
\frac{\sin(\muzero_2c_2)\sin(\muzero_3c_3)}{\muzero_2\muzero_3}p_2p_3+
\text{cyclic terms}
\right\}.\nn
\ea

Again, the equations of motion are given by the Hamilton's equations and
the constraint that the Hamiltonian must vanish:
\ba
\label{eqn:qm0 eom 1}
\frac{dp_\ph}{dt'}&=&\{p_\ph,H'_{\m_o}\}=0\quad \Rightarrow\
p_\ph\ \text{is constant},\\
\label{eqn:qm0 eom 2}
\frac{d\ph}{dt'}&=&\{\ph,H'_{\m_o}\}=p_\ph,
\ea
\ba
\label{eqn:qm0 eom 3}
\frac{dc_1}{dt'}&=&\{c_1,H'_{\m_o}\}=8\p G\g\,\frac{\partial\, H'_{\m_o}}{\partial p_1}\nn\\
&=&-\g^{-1}\,
\frac{\sin(\muzero_1c_1)}{\muzero_1}\nn\\
&&\quad\ \times
\left[\frac{\sin(\muzero_2c_2)}{\muzero_2}p_2
+\frac{\sin(\muzero_3c_3)}{\muzero_3}p_3\right],\\
\label{eqn:qm0 eom 4}
\frac{dp_1}{dt'}&=&\{p_1,H'_{\m_o}\}=-8\p G\g\,\frac{\partial\, H'_{\m_o}}{\partial c_1}\nn\\
&=&\g^{-1}p_1\cos(\muzero_1c_1)\nn\\
&&\quad\ \times
\left[\frac{\sin(\muzero_2c_2)}{\muzero_2}p_2
+\frac{\sin(\muzero_3c_3)}{\muzero_3}p_3\right],
\ea
as well as
\ba\label{eqn:qm0 eom 5}
&&H'_{\m_o}(c_I,p_I)=0\quad
\Rightarrow\qquad p_\ph^2=\\
&&\quad\frac{1}{4\p G\g^2}
\left\{
\frac{\sin(\muzero_2c_2)\sin(\muzero_3c_3)}{\muzero_2\muzero_3}p_2p_3+
\text{cyclic terms}
\right\}.\nn
\ea

From \eqnref{eqn:qm0 eom 3} and \eqnref{eqn:qm0 eom 4}, we have
\be\label{eqn:qm0 dpc/dt'}
\frac{d}{dt'}\left[p_I\frac{\sin(\muzero_Ic_I)}{\muzero_I}\right]=0,
\ee
which gives
\be\label{eqn:qm0 pc}
p_I\frac{\sin(\muzero_Ic_I)}{\muzero_I}
=8\p G\g\hbar\,\K_I.
\ee
Taking \eqnref{eqn:qm0 pc} into \eqnref{eqn:qm0 eom 5}
gives the same constraints on the constant parameters as in \eqnref{eqn:K}.

Substituting \eqnref{eqn:qm0 pc} into \eqnref{eqn:qm0 eom 4} yields
\be\label{eqn:qm0 diff eq 1}
\frac{1}{p_1}\frac{dp_1}{dt'}
=8\p G \hbar\,\cos(\muzero_1c_1)(\K_2+\K_3).
\ee
By \eqnref{eqn:qm0 eom 2} and $\cos x=\pm\sqrt{1-\sin^2x}$,
\eqnref{eqn:qm0 diff eq 1} leads to
\be\label{eqn:qm0 diff eq 2}
\frac{1}{p_I}\frac{dp_I}{d\ph}=
\pm\sqrt{8\p G}\left(\frac{1-\k_I}{\k_\ph}\right)
\left[1-
\left(\frac{\varrho_I}{\varrho^{\m_o}_{I\!,\,{\rm crit}}}\right)^{2/3}\right]^{1/2},
\ee
which gives the bouncing solutions with the behaviors similar to those given by \eqnref{eqn:qm diff eq 2} except that the critical value of $\vr_I$ at which the big bounce takes place is given by
\be
\varrho^{\m_o}_{I\!,\,{\rm crit}}:=
\left[
\left(\frac{\k_\ph}{\k_I}\right)^2\frac{\r_{\rm Pl}\D}{{\muzero_I}^2}
\right]^{3/2}
\frac{1}{p_\ph},
\ee
which can be made arbitrarily small by increasing the value of $p_\ph$. As a result, $\m_o$-scheme gives wrong semiclassical behavior and should be improved by $\mubar$- or $\mubar'$-scheme to fix this problem.

\section{Quantization in the $\mubar$-schemes}\label{sec:mubar schemes}

In this appendix, we give a heuristic argument for the prescription given by \eqnref{eqn:c to sin}, starting from the Hamiltonian constraint of LQG. The motivations for both $\mubar$- and $\mubar'$-schemes are addressed in detail. The advantages and drawbacks of both schemes are also remarked.

The gravitational part of the classical Hamiltonian constraint in the full theory of LQG is given by
\ba\label{eqn:Hamiltonian full}
H_{\rm grav}&=&\frac{1}{8\p G}\int d^3x N e^{-1}
\biggl\{{\e_i}^{jk}F^i_{ab}{\mbox{$\tE$}^a}_j{\mbox{$\tE$}^b}_k\nn\\
&&\qquad\qquad
-2(1+\g^2){K_{[a}}^i{K_{b]}}^j{\mbox{$\tE$}^{a}}_i{\mbox{$\tE$}^{b}}_j\biggr\}\nn\\
&=&-\frac{1}{8\p G\g^2}\int d^3x N e^{-1} {\e_i}^{jk}F^i_{ab}{\mbox{$\tE$}^a}_j{\mbox{$\tE$}^b}_k,
\ea
where $e:=|\det\tilde{E}|^{1/2}$ and in the last line we exploit the fact that for homogeneous and spatially flat models the two terms inside the curly bracket are proportional to each other (because $-\g{K_a}^i={A_a}^i$ as the spatial slice is flat and
$F_{ab}^i={\e^i}_{jk}{A_a}^j{A_b}^j$ as it is homogeneous). Furthermore, the lapse $N$ can be assumed to be constant because of homogeneity and we set $N=1$.

When applied to Bianchi I models, the integral is restricted to a finite sized cell ${\cal V}$ as prescribed by \eqnref{eqn:finite cell}. Accordingly, we should also adopt the replacement rules \eqnref{eqn:replacement rule1} and \eqnref{eqn:replacement rule2}. This procedure gives
\ba
H_{\rm grav}&=&-\frac{1}{8\p G\g^2}\int_{\cal V}d^3x
\,e^{-1}\Vzero^{-2}V^{-2}\nn\\
&&\qquad\times
\left({\e_i}^{jk}{\e^i}_{lm}{A_p}^l{A_q}^m{E^p}_j{E^q}_k\right)\nn\\
&=&-\frac{1}{8\p G\g^2}\int_{\cal V}d^3x
\,e^{-1}\Vzero^{-2}V^{-2}\nn\\
&&\qquad\times
\left(\tilde{c}_2\tilde{c}_3\tilde{p}_2\tilde{p}_3
+\text{cyclic terms}\right)\nn\\
&=&
-\frac{\left(c_2c_3p_2p_3+\text{cyclic terms}\right)}
{8\p G\g^2\,e\,V}
\ea
and
\ba
e^2&=&|\det\tilde{E}|=V^{-3}\abs{\det({E^j}_i\ftriad{a}{j})}\nn\\
&=&V^{-3}\abs{\frac{\tilde{p}_1\tilde{p}_2\tilde{p}_3}
{\azero_1\azero_2\azero_3}}
=V^{-2}\abs{p_1p_2p_3}.
\ea
Consequently, we have
\be\label{eqn:H grav}
H_{\rm grav}=
-\frac{\left(c_2p_2c_3p_3+c_1p_1c_3p_3+c_1p_1c_2p_2\right)}{8\p G\g^2\sqrt{\abs{p_1p_2p_3}}},
\ee
which is exactly the same as $H_{\rm grav}$ given in \eqnref{eqn:cl Hamiltonian}.

When the quantization is performed in the context of LQC, there are two loop quantum corrections. The first is the modification on the inverse triad operator $\widehat{1/\sqrt{\abs{p_I}}}$ as indicated in \eqnref{eqn:f(p_I)}, which is negligible and ignored in this paper.
The second is due to the fact that the connections ${A_a}^i$ (or  $c_I$) do not exist and should be replaced with holonomies (or exponentials of $c_I$).

Following the standard techniques in gauge theories, the curvature component $F^i_{ab}$ can be expressed in terms of holonomies (i.e. Wilson loops). Given a small surface $\a$ center in $\vec{x}$, Stokes' theorem allows us to write the curvature component as
\ba\label{eqn:curvature}
\t_i F^i_{ab}(\vec{x})&\approx&\frac{1}{\e^\a_{ab}}\int_\a \t_i F^i_{cd}\, dx^c\wedge dx^d\nn\\
&\approx&\frac{1}{\e^\a_{ab}}\left[{\cal P}\exp\left(\oint_{\partial\a} \t_i{A_c}^i\, dx^c\right)-1\right],\quad
\ea
where $2i\t_i=\s_i$ are the Pauli matrices; $\partial\a$ is the boundary loop of $\a$; and the \emph{coordinate} area of $\a$ projected in ``$ab$-direction'' is given by
\be
\e^\a_{ab}=\int_{\a}dx^a\wedge dx^b.
\ee
This is a good approximation provided $\e^\a_{ab}$ is small enough and in fact it becomes exact in the continuous limit $\e^\a_{ab}\rightarrow 0$.

For the Bianchi I model, we choose $\a=\Box_{JK}$ to be a small rectangular surface parallel to the $J$-$K$ plane. The \emph{coordinate} lengthes of the edges of $\a$ are denoted as $\mubar_J L_K$ and $\mubar_K L_K$ and accordingly the \emph{coordinate} area is given by $\e^\a_{JK}\equiv\e^\Box_{JK}=\mubar_J\mubar_KL_JL_K$ (see \figref{fig:fig5}).
Applying \eqnref{eqn:replacement rule2} and noting that $F^i_{ab}={\e^i}_{jk}{A_a}^i{A_b}^j$ gives
\be\label{eqn:Fab}
F^i_{ab}\longrightarrow F^I_{JK}:=\e_{IJK}(L_JL_k)^{-1}c_Jc_K,
\ee
we read off from \eqnref{eqn:curvature} that
\be\label{eqn:FJK}
F^I_{JK}
\approx-\frac{2}{\e^\Box_{JK}}{\rm Tr}
\left[\t_I \left(h_{JK}^{(\mubar_J,\mubar_K)}-1\right)
\right],
\ee
where $(I,J,K)$ is any permutation of $(1,2,3)$ and the holonomy along the edges of $\Box_{JK}$ is:
\be
h_{\Box_{\!JK}}^{(\mubar_J,\mubar_K)}:=h_J^{(\mubar_J)}h_K^{(\mubar_K)}
(h_J^{(\mubar_J)})^{-1}(h_K^{(\mubar_K)})^{-1}
\ee
with $h_I^{(\mubar_I)}$ being the holonomy along the individual edge:
\ba
h_I^{(\mubar_I)}&:=&{\cal P}\exp\left(\int_0^{\mubar_I L_I}\!\!\t_i{A_a}^{i} \,dx^a\right)
\longrightarrow
\exp\left(\mubar_I c_I\t_I\right)\nn\\
&=&\cos\left(\frac{\mubar_I c_I}{2}\right)+2\sin\left(\frac{\mubar_I c_I}{2}\right)\,\t_I,
\ea
which follows
\be\label{eqn:trace}
{\rm Tr}\left[\t_I h_{\Box_{\!JK}}^{(\mubar_J,\mubar_K)}\right]
=-\frac{\e_{IJK}}{2}\sin(\mubar_Jc_J)\sin(\mubar_Kc_K).
\ee
Putting \eqnref{eqn:Fab}, \eqnref{eqn:FJK} and \eqnref{eqn:trace} altogether, we then have
\be
c_Jc_K\approx\frac{\sin(\mubar_Jc_J)\sin(\mubar_Kc_K)}{\mubar_J\mubar_K},
\ee
which is exactly the prescription used in \eqnref{eqn:c to sin}.

If $\Box_{JK}$ shrinks to a point, this gives the continuous limit $\mubar_I\rightarrow0$ and we recover the classical Hamiltonian constraint in \eqnref{eqn:H grav}. However, the very feature of LGC is that the continuous limit does not exist and the failure of the limit to exist is intimately related with the underlying quantum geometry of LQG, where eigenvalues of the area operator are \emph{discrete}. In LQC, to implement the discreteness as imprint from the full theory of LQG, we have to set $\mubar_I$ to be finite. The question now is: What values should $\mubar_I$ be set to faithfully reflect the fundamental discreteness of quantum geometry? Two strategies ($\mubar$- and $\mubar'$-schemes) are discussed below.

Since the eigenvalue spectrum of the area operator has an \emph{area gap} $\D$, to impose the discreteness, it is straightforward to set the \emph{physical} area of $\Box_{JK}$ to be $\D$ (depicted in \figref{fig:fig5}.a); i.e.
\ba
a_Ja_K\e^\Box_{JK}&=&a_Ja_K\mubar_J\mubar_KL_JL_K\nn\\
&=&\abs{\e_{IJK}\,p_I}\,\mubar_J\mubar_K=\D,
\ea
which gives the ``$\mubar'$-scheme'' in \eqnref{eqn:mubar'}.

However, a closer consideration seems to suggest that equating the physical area of $\Box_{JK}$ to $\D$ might not be completely sensible. Recall that the validity of \eqnref{eqn:curvature} is based on Stokes' theorem, which does \emph{not} invoke the metric of the surface $\a$ at all. The smallness of $\a$ to ensure good approximation of \eqnref{eqn:curvature} does not directly refer to metric smallness, either.
Furthermore, if we think of the picture of spin networks in LQG, what associated with the physical areas are links (holonomies) of the spin network; the area enclosed by the loop of holonomy links is irrelevant! Therefore, back to LQC, instead of shrinking the area of $\Box_{JK}$, we should associate each \emph{edge} of $\Box_{JK}$ with an area and then shrink the associated areas to $\D$. (See \figref{fig:fig5}.b.)
The edge of $\Box_{JK}$ in $J$-direction is of \emph{coordinate} length $\mubar_JL_J$, with which, most naturally, we associate a rectangle $\boxdot_J$ parallel to $I$-$K$ plane of \emph{coordinate} lengths $\mubar_JL_I$ and $\mubar_JL_K$ on its edges. (That is, the ratio of the edge of $\boxdot_J$ to the edge of ${\cal V}$ in the $I$/$K$-direction is set to be the same as the ratio of the edge of $\Box_{JK}$ to the edge of ${\cal V}$ in the $J$-direction.) We then set the \emph{physical} area of $\boxdot_J$ to $\D$; i.e.
\be
a_Ia_K\mubar_J^2L_IL_K=\mubar_J^2\abs{p_J}=\D,
\ee
which gives the ``$\mubar$-scheme'' in \eqnref{eqn:mubar}.

[Note that in both pictures of $\mubar'$- and $\mubar$-schemes, if we rather set the \emph{coordinate} area of $\Box_{JK}$ or $\boxdot_J$ to be $\D$, we end up with constant $\mubar_I$. This is the ``$\m_o$-scheme'' used in the precursor strategy. The wrong semiclassical behavior of the $\m_o$-scheme originates from the problem that the \emph{coordinate} area (both of $\Box_{JK}$ and of $\boxdot_{J}$) is not totally physically relevant.]

In the full theory of LQG, when the Hamiltonian acts on an spin network state, it adds a new link of spin-$1/2$ and the coloring of the links on which the new link is attached is increased or decreased by $1/2$ (see \figref{fig:fig6}.a-b). The resulting spin network state is equivalent to the original state superimposed with a triangular loop of spin-$1/2$ links (\figref{fig:fig6}.c). This triangular loop is essentially the Wilson loop discussed above. From this perspective, the $\mubar'$-scheme can be understood as associating the area in pink in \figref{fig:fig6}.d with $\D$, whereas the $\mubar$-scheme as associating the area in blue in \figref{fig:fig6}.e with $\D$. In the context of spin network states, the coloring of a link corresponds to the area of the surface to which the link penetrates and the smallest coloring spin-$1/2$ gives rise to the \emph{area gap} $\D$. Therefore, it is in this sense that $\mubar$-scheme is a more direct implementation of the underlying discreteness of quantum geometry than $\mubar'$-scheme.

Furthermore, in the full quantum theory of LQC, $\mubar$-scheme has the important virtue that we can define the affine variables $v_I$ via
\be
\frac{\partial}{\partial v_I}:=4\p\g\Pl^2\,\mubar_I\frac{\partial}{\partial p_I}
\ee
such that Hamiltonian constraint of the full quantum theory gives the evolution as a difference equation in terms of $v_I$ and therefore the methodology used for the isotropic LQC can be easily applied \cite{Chiou:2006qq}. This strategy fails in $\mubar'$-scheme since
\be
\left[\mubar'_I\frac{\partial}{\partial p_I},\ \mubar'_J\frac{\partial}{\partial p_J}\right]\neq 0
\ee
and hence the corresponding affine variables do not exist. This makes it difficult to construct the full quantum theory of Bianchi I LQC in $\mubar'$-scheme.

On the other hand, as studied in \secref{sec:effective dynamics}, $\mubar'$-scheme has the advantage over $\mubar$-scheme that the effective dynamics in $\mubar'$-scheme is independent of the choice of ${\cal V}$. The difference for this point between $\mubar$- and $\mubar'$-schemes can be understood, heuristically but instructively, by estimating the quantity $\mubar_Ic_I$ with the classical formulae; that is, plugging \eqnref{eqn:c and HI} for $c_I$, we have
\ba
\label{eqn:mubar c}
\mubar_1c_1&\approx&\g\D^{1/2}H_1\left(\frac{{\bf L}_1^2}{{\bf L}_2{\bf L}_3}\right)^{1/2},\\
\label{eqn:mubar' c}
\mubar'_Ic_I&\approx&\g\D^{1/2}H_I.
\ea
Since the quantity $\mubar_Ic_I$ indicates how significant the quantum corrections are (quantum corrections are negligible if $\mubar_Ic_I\ll 1$ ), \eqnref{eqn:mubar c} tells that in $\mubar$-scheme, the place at which the quantum effects become appreciable is tied up with not only the Hubble rates $H_I$ but also the physical geometry of ${\cal V}$. By contrast, \eqnref{eqn:mubar' c} shows that the different choice of ${\cal V}$ is irrelevant in $\mubar'$-scheme.

In the language of \cite{Bojowald:2007ra}, $\mubar$-scheme corresponds to a lattice refinement model whereby the number of vertices is proportional to the transverse area. A stability analysis suggests that $\mubar$-scheme leads to an unstable difference equation in the full theory of LQC and thus may not represent a good quantization scheme.

Both $\mubar$- and $\mubar'$-schemes have desirable and undesirable features of their own. In order to understood them more deeply, in the main text we study both schemes and their ramifications at the level of effective dynamics.

\begin{widetext}

\begin{figure}
\begin{picture}(510,90)(0,0)

\put(160,80){(a)}
\put(350,80){(b)}

\put(40,5){
\scalebox{0.6}
{\includegraphics{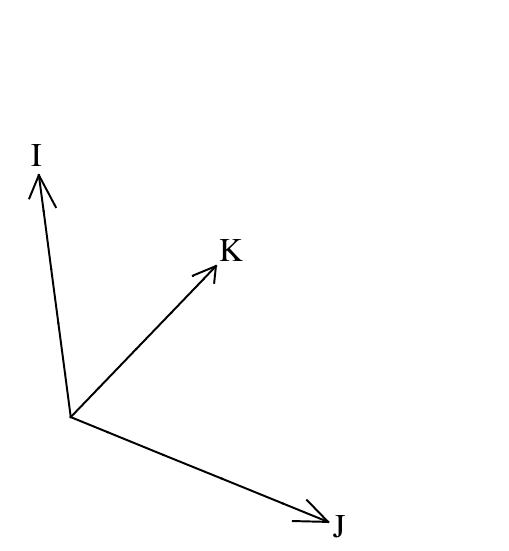}}
}

\put(140,-32){
\scalebox{0.70}
{\includegraphics{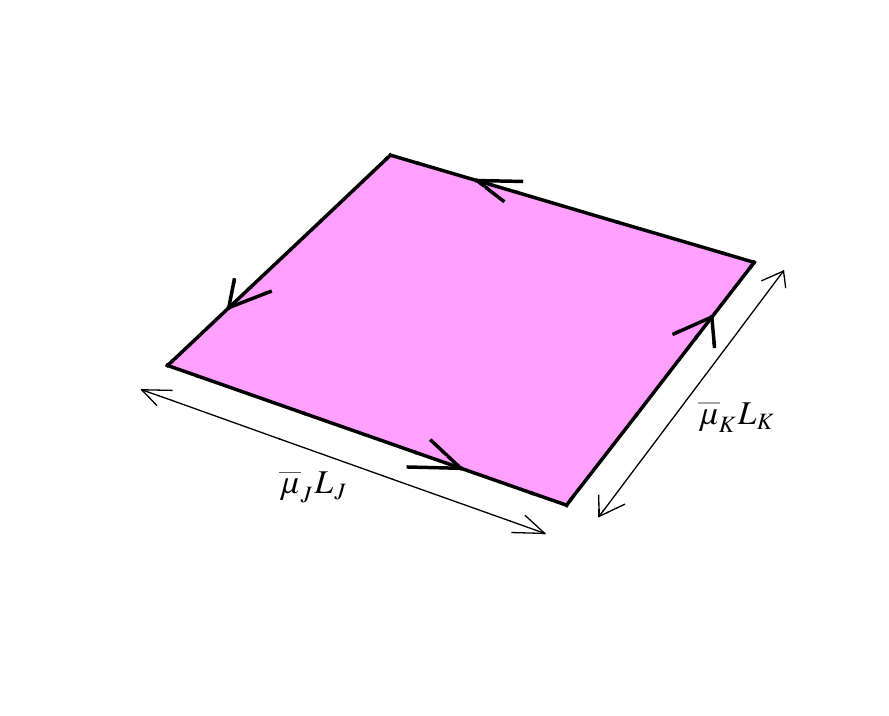}}
}

\put(330,-32){
\scalebox{0.70}
{\includegraphics{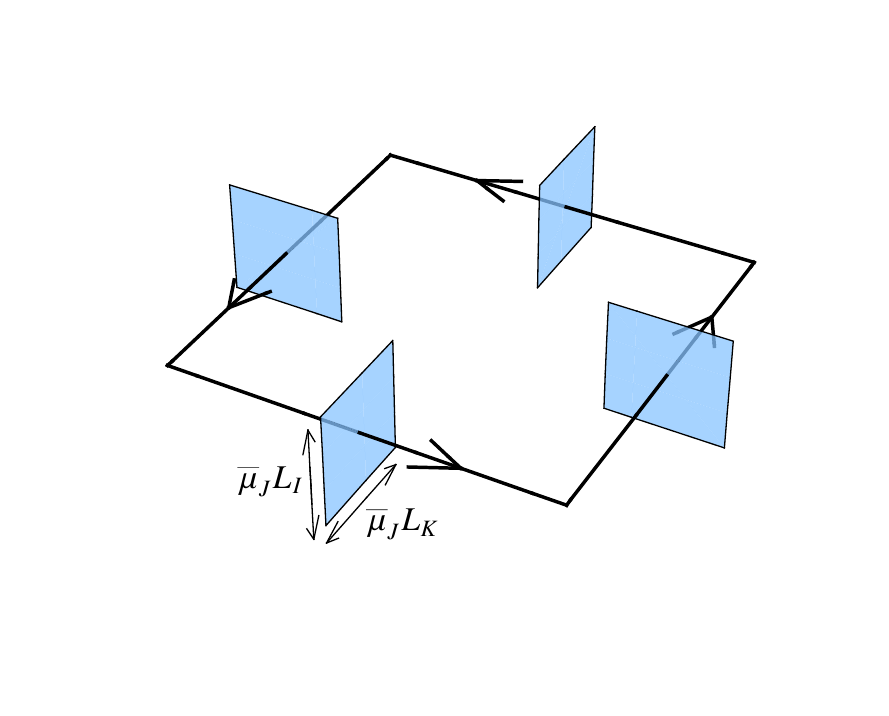}}
}

\end{picture}
\caption{\textbf{(a)} The surface in pink is $\Box_{JK}$, the \emph{physical} area of which is to be shrunk to $\D$ in the $\mubar'$-scheme. \textbf{(b)} The surfaces in blue are $\boxdot_J$ and $\boxdot_K$, the \emph{physical} areas of which are to be shrunk to $\D$ in the $\mubar$-scheme.}\label{fig:fig5}
\end{figure}

\begin{figure}
\begin{picture}(510,100)(0,0)

\put(70,50){$\longrightarrow$}
\put(180,50){$=$}
\put(280,50){$\Longrightarrow$}
\put(390,50){or}
\put(5,90){(a)}
\put(100,90){(b)}
\put(200,90){(c)}
\put(310,90){(d)}
\put(415,90){(e)}

\put(-15,-30){
\scalebox{0.45}
{\includegraphics{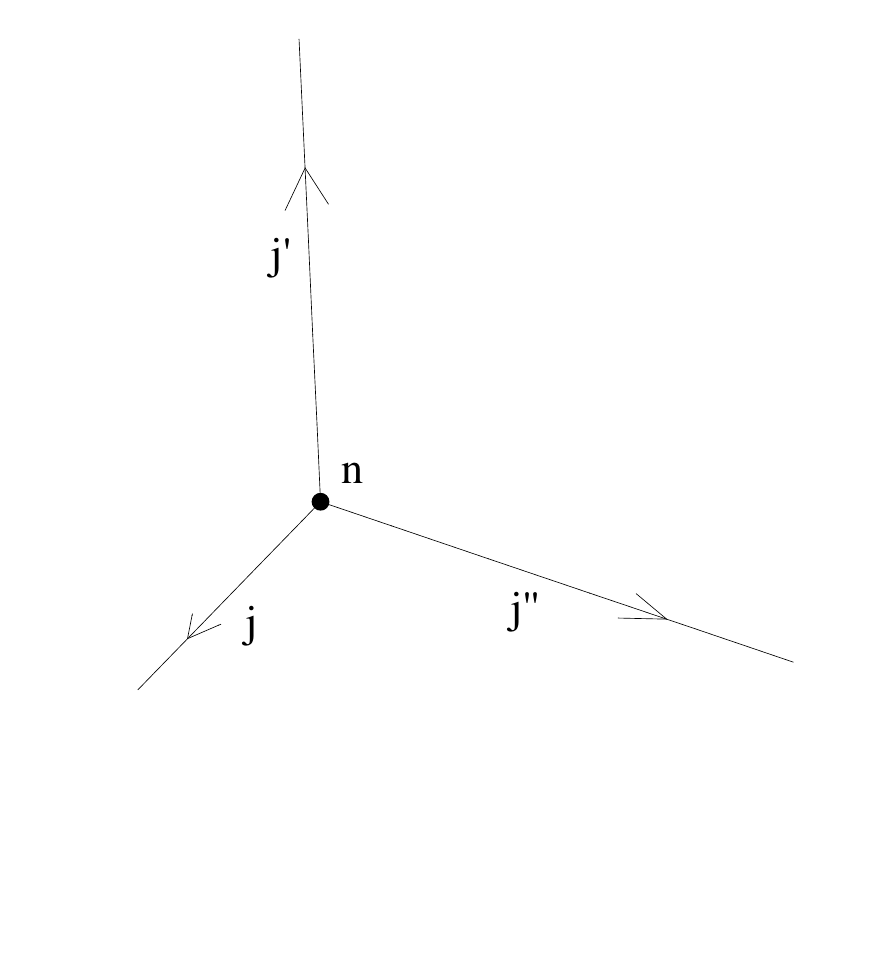}}
}

\put(85,-30){
\scalebox{0.45}
{\includegraphics{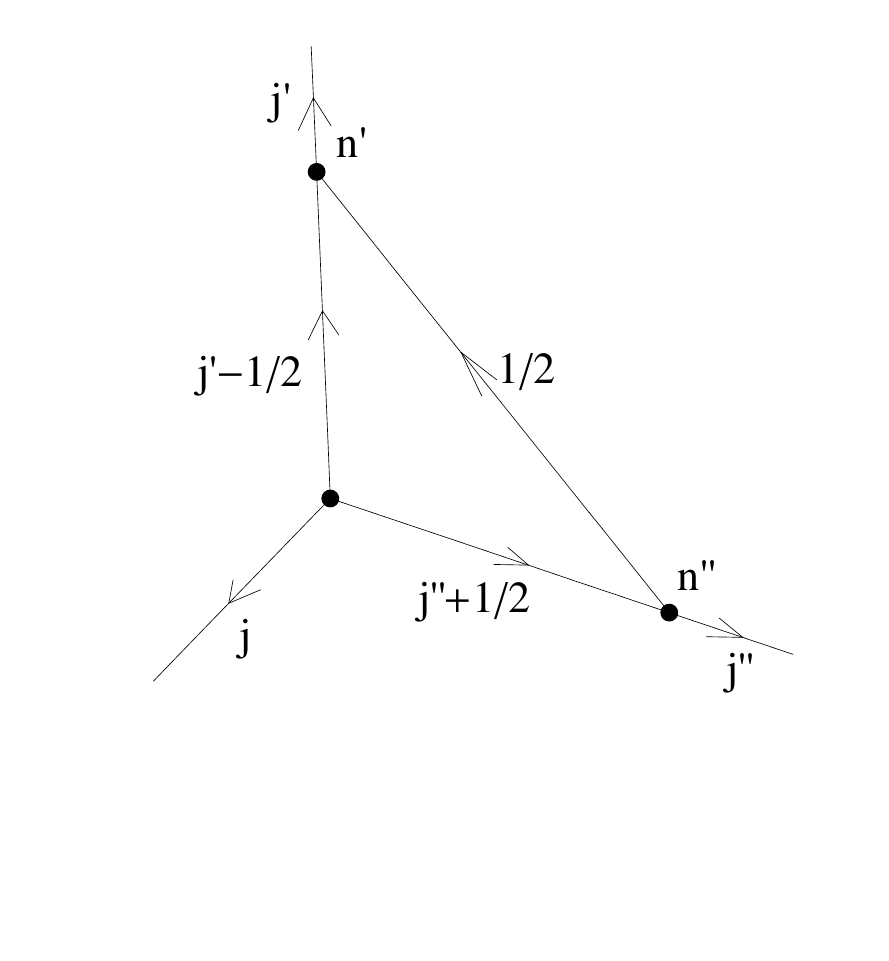}}
}

\put(190,-30){
\scalebox{0.45}
{\includegraphics{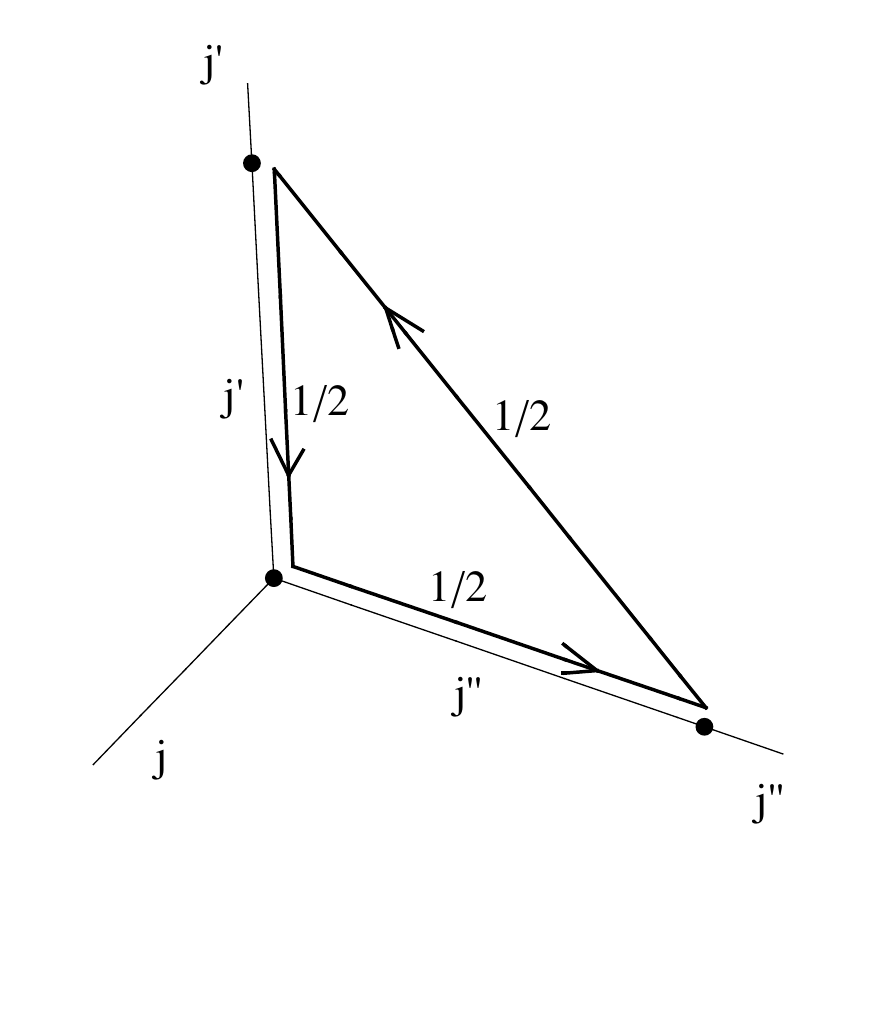}}
}

\put(300,-30){
\scalebox{0.45}
{\includegraphics{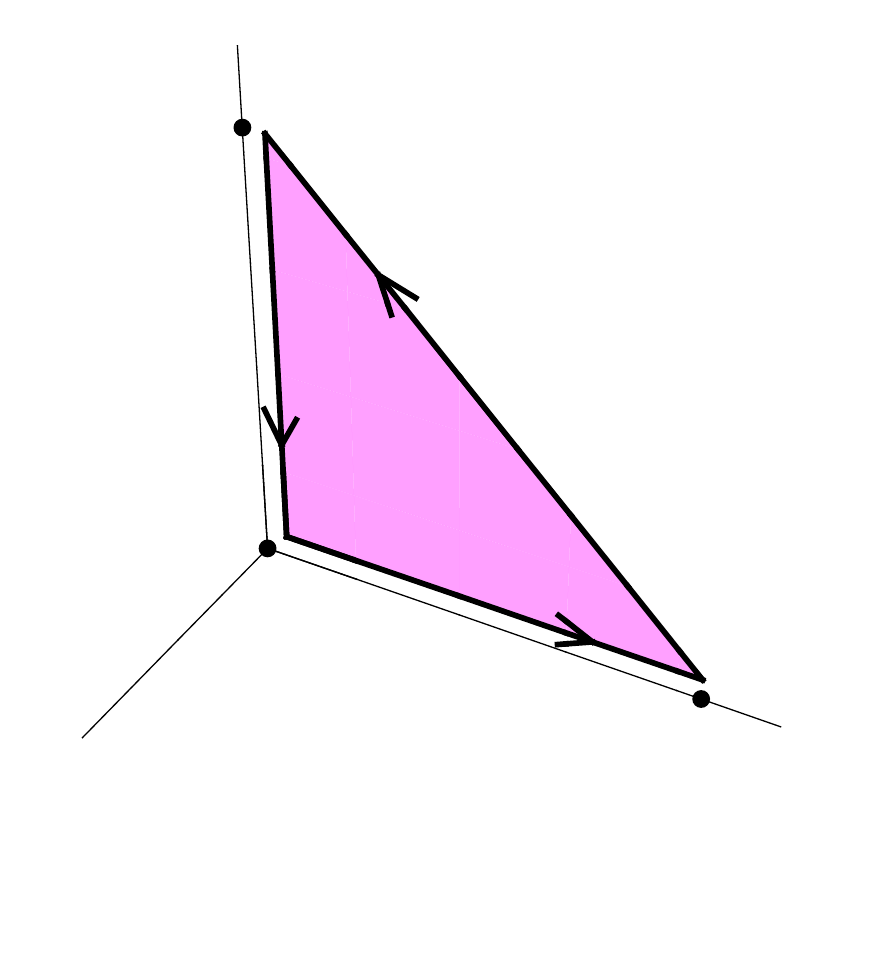}}
}

\put(400,-30){
\scalebox{0.45}
{\includegraphics{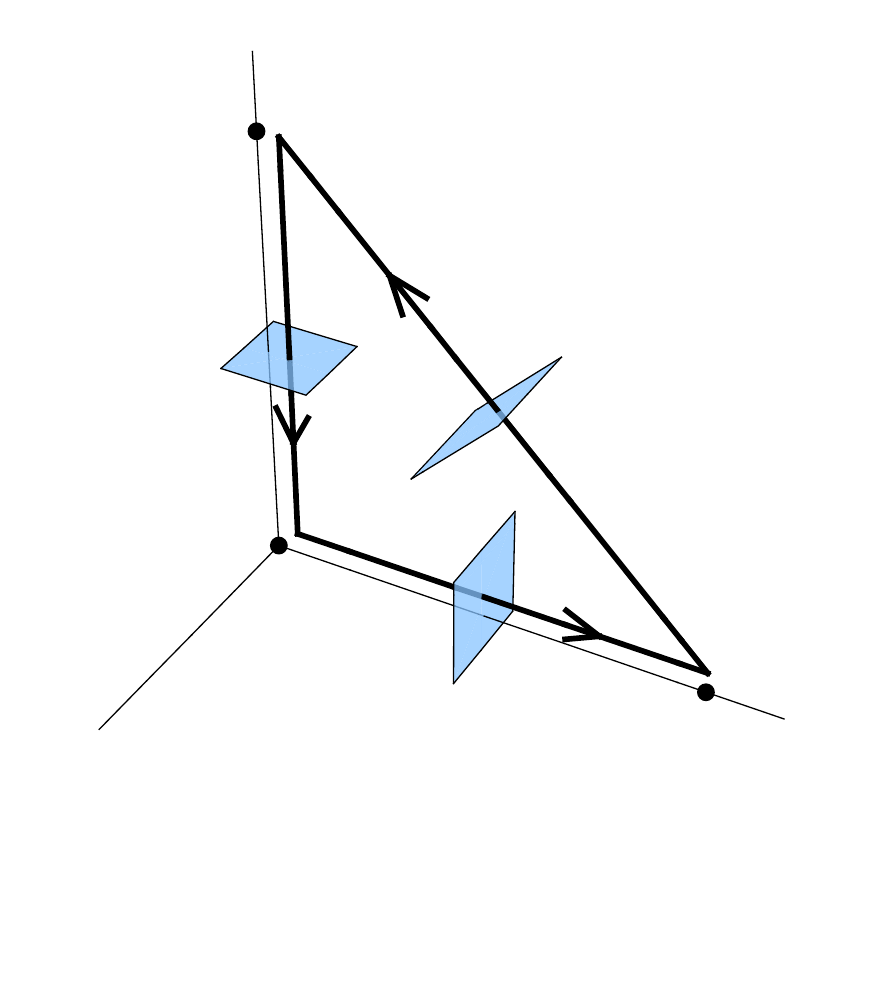}}
}

\end{picture}
\caption{\textbf{(a)-(b)} The action of Hamiltonian acting on a spin network state. \textbf{(c)} The resulting state is equivalent to the original one superposed with a triangular loop. \textbf{(d)} The idea of $\mubar'$-scheme. \textbf{(e)} The idea of $\mubar$-scheme.}\label{fig:fig6}
\end{figure}

\end{widetext}

\end{document}